\documentclass[12pt]{article}
\usepackage{a4wide}
\usepackage{latexsym}
\usepackage{amsfonts}
\usepackage{amssymb}
\usepackage{axodraw}

\def\l{\langle}
\def\r{\rangle} 
\def\bq{\begin{eqnarray}}
\def\eq{\end{eqnarray}}
\def\v{\verb}
\def\bs{\begin{small}}
\def\es{\end{small}}
\def\eps{\varepsilon}

\newcounter{exercise}
\newcounter{example}
\begin{document}

\thispagestyle{empty}

\begin{flushright}
UPRF-02-015
\end{flushright}
\vspace{1.5cm}

\begin{center}
  {\Large \bf Computer Algebra in Particle Physics}\\[.3cm]
  \vspace{1.7cm}
  {\sc Stefan Weinzierl$^{1}$}\\
  \vspace{1cm}
  {\it Dipartimento di Fisica, Universit\`a di Parma,\\
       INFN Gruppo Collegato di Parma, 43100 Parma, Italy} \\
\end{center}

\vspace{2cm}

\begin{abstract}\noindent
  {These lectures given to graduate students in theoretical particle physics,
   provide an introduction to the ``inner workings'' of computer algebra
   systems.
   Computer algebra has become an indispensable tool for 
   precision calculations in particle physics. 
   A good knowledge of the basics of computer algebra systems allows one
   to exploit these systems more efficiently.
  }
\end{abstract}

\vspace*{\fill}

 \noindent 
 $^1${\small email address : stefanw@fis.unipr.it}

\newpage
\tableofcontents
\newpage

\reversemarginpar

\section{Introduction}

Many of the outstanding calculations leading to precise predictions
in elementary particle physics would not have been possible
without the help of computer algebra systems (CAS).
Examples of such calculations are 
the calculation of the anomalous magnetic moment of the electron 
up to three loops \cite{Laporta:1996mq},
the calculation of the ratio
\bq
R & = & \frac{e^+ e^- \rightarrow \mbox{hadrons}}
             {e^+ e^- \rightarrow \mu^+ \mu^-}
\eq
of the total cross section for hadron production to the total
cross section for the production of a $\mu^+ \mu^-$ pair
in electron-positron annihilation to order $O\left( \alpha_s^3 \right)$
\cite{Gorishnii:1991vf} (also involving a three loop calculation)
or the calculation of the 4-loop contribution to the 
QCD $\beta$-function \cite{vanRitbergen:1997va}.
While the examples cited above have been carried out to an impressive
third or fourth order, they only involve a single scale.
Additional complications occur if multiple scales are involved.
This is a challenging problem and computational quantum field theory 
witnessed a tremendous boost in the past two years in this area.
The breakthrough was the development of
systematic algorithms, which allowed the calculation of higher loop amplitudes
with more than one scale.
Examples are here the calculation of the most interesting
$2 \rightarrow 2$ processes 
\cite{Bern:2000dn,Bern:2000ie,Anastasiou:2000kg,Binoth:2002xg}, 
which depend on two scales
and the calculation for $e^+ e^- \rightarrow \mbox{3 jets}$
\cite{Garland:2001tf,Moch:2002hm}, 
which depends on three scales.
None of the examples cited above could have been done without the help of 
computer algebra.
Turning the argument around, these calculations are just the ones
which are at the edge of what is feasible with todays methods and technology.
Limiting factors are not only missing knowledge how to calculate certain processes,
but also the speed and memory of computers, 
missing knowledge on efficient algorithms or improper
representation of the data inside the computer.
Understanding the basics of computer algebra systems
allows one to exploit these systems more efficiently.
This is the main goal of these lectures.
\\
\\
Particle physics is one important field of application for computer algebra
and exploits the capabilities of computer algebra systems.
This leads to valuable feed-back for the development of computer algebra systems.
Quite a few computer algebra systems have their roots within the high energy
physics community or strong links with them:
REDUCE \cite{REDUCE}, 
SCHOONSHIP \cite{SCHOONship}, 
MATHEMATICA \cite{Mathematica},
FORM 
\cite{FORM}
or GiNaC \cite{Bauer:2000cp}, 
to name only a few.
Looking at the history of computer algebra systems, the first programs date back to the 1960's.
Fig. (\ref{survey}) gives a historical overview together with the first appearance of some
programming languages.
The first systems were almost entirely based on LISP (``LISt Programming language'').
LISP is an interpreted language and, as the name already indicates, designed for the
manipulation of lists.
Its importance for symbolic computer programs in the early days can be compared to
the importance of FORTRAN for numerical programs in the same period.
Already in this first period, the program REDUCE had some special features for the application
to high energy physics.
An exception to the LISP-based programs was SCHOONSHIP, written in assembler language by Veltman
and specially designed for applications in particle physics.
The use of assembler code lead to an incredible 
fast program (compared to the interpreted programs at that time)
and allowed the calculation of more complex scattering processes in high energy physics.
The importance of this program was recognized in 1998 by awarding the Nobel prize
to M. Veltman.
Also the program MACSYMA \cite{Maxima}
deserves to be mentioned explicitly, since it triggered 
important development with regard to algorithms.
\begin{figure}
\begin{center}
\begin{tabular}{|rcl|}
\hline
 & & \\
& \multicolumn{2}{l}{\it The early days, mainly LISP based systems} \vline \\
 & & \\
1958 & FORTRAN     & \\
1960 & LISP        & \\
1965 & & MATHLAB     \\
1967 & & SCHOONSHIP  \\
1968 & & REDUCE      \\
1970 & & SCRATCHPAD, evolved into AXIOM  \\
1971 & & MACSYMA     \\
1979 & & muMATH, evolved into DERIVE \\
 & & \\
& \multicolumn{2}{l}{\it Commercialization and migration to C} \vline \\
 & & \\
1972 & C & \\
1981 & & SMP, with successor MATHEMATICA \\
1988 & & MAPLE       \\
1992 & & MuPAD \\
 & & \\
& \multicolumn{2}{l}{\it Specialized systems} \vline \\
 & & \\
1975 & & CAYLEY (group theory), with successor MAGMA\\
1985 & & PARI (number theory calculations) \\
1989 & & FORM (particle physics)\\
1992 & & MACAULAY (algebraic geometry) \\
 & & \\
& \multicolumn{2}{l}{\it A move to object-oriented design and open-source} \vline \\
 & & \\
1984 & C++ & \\
1995 & Java & \\
1999 & & GiNaC \\
& & \\
\hline
\end{tabular}
\caption{\label{survey} \it A historical survey on computer algebra system 
together with the first appearance of some programming languages.}
\end{center}
\end{figure}
In the 1980's new computer algebra systems started to be written in C. This allowed
to exploit better the resources of the computer (compared to the interpreted language LISP)
and at the same time allowed to maintain portability (which would not have been 
possible in assembler
language).
This period marked also the appearence of the first commercial computer algebra system,
among which Mathematica and Maple \cite{Maple} are the best known examples.
In addition, also a few dedicated programs appeared, an example relevant to particle
physics is the program FORM by J. Vermaseren as a (portable) successor to SCHOONSHIP.
In the last few years issues of the maintainability of large projects became
more and more important and the overall programming paradigma changed from
procedural programming to object-oriented design.
In terms of programming languages this was reflected by a move from C to C++.
Following this change of paradigma, the library GiNaC was developed.
The GiNac library allows symbolic calculations in C++.
At the same time the last few years marked also a transition away from commercial systems
(back) to open-source projects.
\\
\\
The choice of the approriate computer algebra system depends on the personal needs.
It is worth analyzing the problem to be solved first.
One class of problems requires only basic support from the computer algebra systems.
These problems are often ``local'' problems, where the problem expands into a sum of different terms and
each term can be solved independently of the others.
The complication which occurs is given by the fact that the number of terms can become
quite large.
Here the requirements on a computer algebra system are bookkeeping 
and the ability to handle large amounts of data.
Quite a few problems in high energy physics fall into this class.
A notorious example are tree-level processes involving several
triple gauge-boson vertices.
The Feynman rules, like the one for the three-gluon vertex, 
\begin{eqnarray}
\begin{picture}(50,30)(0,15)
 \Gluon(0,20)(30,20){4}{2}
 \Vertex(30,20){2}
 \Gluon(30,20)(50,40){4}{2}
 \Gluon(50,0)(30,20){4}{2}
\end{picture} & = &
 g f^{abc} \left[
       g^{\nu \lambda} \left( k_3^\mu - k_2^\nu \right)
      +g^{\lambda \mu} \left( k_1^\nu - k_3^\nu \right)
      +g^{\mu \nu} \left( k_2^\lambda - k_1^\lambda \right)
     \right], \\
& & \nonumber 
\end{eqnarray}
expand this vertex into up to six terms, which easily blow up the size
of intermediate expressions.
\\
The second class of problems requires more sophisticated methods.
These can be either standarized non-local operations (factorization is an example),
which to some extend are already implemented in some computer algebra systems,
or dedicated algorithms, which are developed by the user to solve this particular
problem.
Here the ability to model abstract mathematical concepts in the programming language of the computer algebra
system is essential.
\\
Certainly, the two complications can also occur at the same time, e.g. the need to implement
dedicated algorithms and the need to process large amounts of data.
Higher loop calculations in quantum field theory tend to fall into this category.
To summarize, requirements on a computer algebra system are therefore:
\begin{itemize}
\item Efficiency in performance: If the system needs to process 
large amounts of data, 
performance is a priority.
Usually this implies that the user's program is compiled and not interpreted.
Systems optimized for performance may also contain low-level routines, which
exploit efficiently the resources of the computer (memory, hard disc).
\item Support of object-oriented programming: This is of importance for users,
who would like to implement complex algorithms for abstract mathematical entities.
Being able to program at an abstract level (e.g. not in a low-level computer language)
can reduce significantly the number of bugs.
In a similar direction goes the requirement for strong type checking.
Strong type checking can catch a number of bugs already at compilation time.
\item Short development cycles: It is usually the case that most time is spent
for the development and the implementation of algorithms. The actual running time
of the completed program is usually negligible against the development time.
Since computer programs are developed by humans,
the development time can be reduced if the computer algebra
system allows for interactive use and provides high-quality output on the screen.
\item Source code freely available: No computer program is free of bugs, neither commercial
products nor non-commercial programs.
The probability of finding a bug in a particular computer algebra system
while working on a large project with this computer algebra system
is not negligible.
If the source code is available it is simpler to trace down the bug, understand
its implications and to fix it.
\end{itemize}
Unfortunately, there is no system which would fullfill all requirements and the choice
of the appropriate system involves therefore some inevitable trade-off.
As examples for computer algebra systems I summarize the main features of 
Maple, FORM and GiNaC:
\begin{itemize}
\item Maple: This is a commercial product. The advantages are a graphical user interface and the
support for some sophisticated operations like factorization, integration or summation.
The disadvantages are that Maple is quite inappropriate when it comes to processing large amounts of data.
\item FORM: The advantage of FORM is its speed and its capability to handle large amounts of data.
It is widely used within the high energy community and the program of choice
for calculations involving large intermediate expressions.
\item GiNaC: This is a library for symbolic computations in C++. 
The main feature is the possibility to implement one's
own algorithms in an object-oriented way.
It can handle large amounts of data and for some benchmark tests 
the speed is comparable to FORM.
\end{itemize}
These lecture notes follow a bottom-up approach from basic
data structures to complex algorithms for loop calculations in high
energy physics.
In the next section I discuss data structures, object-oriented programming
and the design of a simple toy computer algebra system.
The discussion of the design will follow closely the design of the 
GiNaC-library.
This is partly motivated by the fact that here the source code is freely
available, as well as partly by the fact that I am involved in 
the development of 
a program \cite{gtybalt}, which combines the GiNaC-library with
a C/C++ interpreter and a what-you-see-is-what-you-get editor.
The resulting system allows an interactive use and displays results
in metafont-type quality.
\\
Section 3 deals entirely with efficiency. The efficient implementation of
algorithms is discussed in detail for three examples:
Shuffling relations, the multiplication of large numbers and the
calculation of the greatest common denominator.
\\
The development of computer algebra systems initiated also research on
systematic algorithms for the solution of certain mathematical problems.
In section 4 I discuss some of the most prominent ones:
Factorization, symbolic integration, symbolic summation and simplifications
with the help of Gr\"obner bases.
\\
Section 5 is devoted to computational perturbative quantum field theory.
I review algorithms, which have been developed recently,
for the calculation of multi-loop integrals.
\\
General textbooks, on which parts of these lectures are based, are:
D. Knuth, ``The Art of Computer Programming'' \cite{Knuth},
K. Geddes et al., ``Algorithms for Computer Algebra'' \cite{Geddes} and
J. von zur Gathen and J. Gerhard, ``Modern Computer Algebra'' \cite{Gathen}.

\section{Data structures}

In this section I start from the basics: How data structures
representing mathematical expressions, can be stored in the physical
memory of the computer.
Additional information on this subject can be found in 
the excellent book by Knuth \cite{Knuth}.
The first programs for symbolic computations implemented mathematical
expressions by nested lists and I discuss as an example symbolic 
differentiation in LISP first.
Lists are special cases of ``containers'', and alternative types
of containers are discussed in the following.
In particular I examine the CPU time necessary to access or operate
on the elements in various containers.
The size of programs for the solution of certain problems depends
on the complexity of the problem and can become large.
In order to avoid that these programs become unmanageable,
object-oriented techniques have been invented.
I review these techniques briefly.
At the end of this section I discuss the design of a simple
toy computer algebra system.

\subsection{Lists}

I start the section on data structures with a concrete example:
Symbolic differentiation programmed in LISP. 
This is THE classic example for symbolic calculations.
Symbolic differentiation can be specified by a few rules:
\bq
\frac{d}{dx} c & = & 0, \nonumber \\
\frac{d}{dx} x & = & 1, \nonumber \\
\frac{d}{dx} \left( f(x) + g(x) \right) & = & 
 \frac{d}{dx} f(x) + \frac{d}{dx} g(x), \nonumber \\
\frac{d}{dx} \left( f(x) g(x) \right) & = &
 \left( \frac{d}{dx} f(x) \right) g(x) + f(x) \left( \frac{d}{dx} g(x) \right).
\eq
These rules are sufficient to differentiate polynomials in $x$.
An implementation in 
LISP\footnote{For the nostalgic: A free LISP interpreter is available from
http://www.gnu.org/software/gcl/gcl.html.
A good introduction to LISP and programming techniques in LISP can be found in
\cite{Winston}.}
looks like this:
\begin{verbatim}
(DEFUN OPERATOR (LIST) (CAR LIST))

(DEFUN ARG1 (LIST) (CADR LIST))

(DEFUN ARG2 (LIST) (CADDR LIST))

(DEFUN DIFF (E X)
  (COND ((ATOM E) (COND ((EQUAL E X) 1)
                        (T 0)))
        ((EQUAL (OPERATOR E) '+)
         `(+ ,(DIFF (ARG1 E) X) ,(DIFF (ARG2 E) X)))
        ((EQUAL (OPERATOR E) '*)
         `(+ (* ,(DIFF (ARG1 E) X) ,(ARG2 E))
             (* ,(ARG1 E) ,(DIFF (ARG2 E) X))))))
\end{verbatim}
A few comments on this short program are in order.
Within LISP the prefix notation is usually used, e.g. $A+B$ is represented
by \v/(+ A B)/.
The first three lines define aliases to the build-in LISP functions 
\v/CAR/, \v/CADR/ and \v/CADDR/ to extract
the first, second or third element of a list, respectively.
They are only introduced to make the program more readable.
The program itself shoud be readable even without the knowledge of LISP,
except for the appearance of the single quote `` ' '', the backquote `` ` ''
and the comma ``,'' character.
LISP generally interprets the first element in a list as the name of an
operation and tries to evaluate this operation with the remaining elements
of the list as arguments.
This behaviour can be prohibited by putting a single quote `` ' '' in front
of the list and the list remains unevaluated.
The backquote `` ` '' acts like the single quote, except that any commas
that appear within the scope of the backquote have the effect
of unquoting the following expression.
\\
In this simple example it is further assumed that addition and multiplication
take only two arguments.
LISP is an interactive language and entering
\begin{verbatim}
                         (DIFF '(* A X) 'X)
\end{verbatim}
at the prompt for 
\bq
\frac{d}{dx} \left( a x \right)
\eq
yields
\begin{verbatim}
                         (+ (* 0 X) (* A 1)),
\end{verbatim}
which stands for
\bq
(0 \cdot x ) + (a \cdot 1).
\eq
Simplifications like $0 \cdot x = 0$, $a \cdot 1 = a $ or $0+a=a$ are out of the
scope of this simple example.
This example already shows a few important features of computer algebra programs:
\begin{itemize}
\item The distinction between objects, which contain sub-objects and objects without
further substructures. The former are generally referred to as ``containers'', the later
are called ``atoms''. Examples for atoms are symbols like $a$ or $x$, while examples
for containers are data structures which represent multiplication or addition of some
arguments.
\item The use of recursive techniques. In the example above the function \v/DIFF/ calls
itself whenever it encounters a multiplication or addition. Note that the recursive
function call has simpler arguments, so that the recursion will terminate.
\item Lists are used to represent data structures like addition and multiplication.
The first element in the list specifies what the list represents.
\item Lists can be nested. The output of the example above, \v/(+ (* 0 X) (* A 1))/,
consists of two lists nested inside another list.
\item The output of a symbolic procedure is not necessarily in the most
compact form.
\end{itemize}

\subsection{Containers}

A container is an object that holds other objects. Lists and arrays are examples
of containers.
There are several ways how the information of a container can be stored in physical
memory.
In particular, the time needed to access one specific element will depend on the
lay-out of the data in the memory.
The appropriate choice depends on the specific problem under consideration.
In this context, the ``big-O'' notation is useful: 
An indication $O(n)$ for an operation on $n$ elements
means that the operation takes time proportional to the number of
elements involved.
\begin{figure}
\begin{center}
\begin{tabular}{|ll|}
\hline
 $O(1)$ & cheap \\
 $O( \log(n) )$ & fairly cheap \\
 $O(n)$ & expensive \\
 $O(n \log(n) )$ & expensive \\
 $O(n^2)$ & very expensive \\
\hline
\end{tabular}
\caption{\label{Onotation} \it A summary on the costs of certain operations.
Operations of order $O(1)$ or $O(\log(n))$ are considered ``cheap'' operations.
Further, $O(n\log(n))$ is considered closer to $O(n)$ than to $O(n^2)$.
It is usually a considerable speep-up, if an operation which
naively takes $O(n^2)$ time, can be improved to $O(n\log(n))$.
In generally one tries to avoid operations, which take $O(n^2)$ time.}
\end{center}
\end{figure}
Fig. (\ref{Onotation}) gives a rule-of-thumb for the significance of the
cost of certain operations.
\\
An array (also called ``vector'' within the C++ terminology) 
stores the information linearly in memory.
Given the fixed size $l$ of one entry and the address $a_1$ of the first entry,
the address $a_i$ of the $i$-th entry is easily obtained as
\bq
a_i & = & a_1 + (i-1) l.
\eq
This involves one multiplication and one addition.
The CPU time to access one specific element is therefore independent of the number
of elements in the array.
However, suppose that we have an array of $n$ elements and that we would like to insert
a new element between the $i$th and $(i+1)$th entry. This involves to shift the entries
with index $n$, $n-1$, $n-2$, ..., $i+1$ by one position and is an $O(n)$ operation.
If these operations occur frequently, an array is not the best suited data structure.
\\
In that case a list structure is more appropriate.
A list is often implemented as a double-linked list, where each node
of the double-linked list contains one field of information, one pointer
to the next element and one pointer to the previous element.
Inserting a new element somewhere in the middle of the list involves only updating
the pointers and is an $O(1)$ operation.
Typical list operations are to add or to remove elements in the middle of the list.
However there is also a drawback for lists. To access the $i$th element 
of a list, there is no other way
than to start at the first element and to follow the pointers to the next elements sequentially, until
one arrives at the $i$th entry. This is an $O(n)$ operation.
\\
A generalization of a double-linked list is a rooted tree, where
each node may have several sub-nodes.
A list can be viewed as a tree, where each node has exactly one sub-node.
An important special case is a binary tree, where each node has up to two sub-nodes.
A binary tree can be used to encode an order relation: Elements of the left sub-tree of a particular
node are ``less-than'' the current node, whereas elements of the right sub-tree are
``greater-than'' the current node.
\\
An associative array (sometimes also called a map) keeps pairs of values.
Given one value, called the key, one can access the other, called the mapped value.
One can think of an associative array as an array for which the index need not be an 
integer.
An associative array can be implemented by a binary tree.
In this case it is further assumed that there is
a less-than operation for the keys and the associative array
keeps it's elements ordered with respect to this relation.
To find the mapped value corresponding to a specific key $k$, one first compares
the key $k$ with the key $k_R$ of the entry at the root of  the tree.
If $k < k_R$, the mapped value is in the left sub-tree
and the procedure is repeated with the top-node of the left sub-tree as root, 
if $k> k_r$ the mapped value is
in the right sub-tree and if $k=k_R$ we have already found the corresponding
(key,value) pair.
To find the mapped value takes on the average $O(\log(n))$ operations.
It is important to note that insertion of new elements is also an
$O(\log(n))$ operation and does not require to sort the complete tree.
Insertion is done as follows: One starts at the root and one compares the 
key $k$ of the new element with the key $k_R$ of the root.
Let us assume that $k>k_R$. If the root has a right sub-tree, the procedure
is repeated with the top-node of the right sub-tree as root.
If the root does not have a right sub-tree, we attach a new right sub-tree
to the root, consisting of the new element.
\\
An alternative implementation for an associative array uses a hash map.
Suppose that there is an easy to calculate function, called hash function,
which maps each key to a specific address in a reserved memory area.
In general, there will be no hash function, which can guarantee that two different
keys are not mapped to the same address, but a good hash function will avoid
these collisions as much as possible.
If a collision occurs, there are several strategies to deal with this case:
The simplest implementation for insertions would just use the next free entry,
whereas to access elements one would do a linear search.
An associative array based on a hash map requires therefore a hash function
and an ``is equal''-operation for its keys.
Since the calculation of the hash value should be a fast operation, many hash functions
are based on exclusive-or operations and bit-wise rotations.
If collisions are not frequent, insertions and access can be done in constant time.
Collisions are less frequent, if the memory area for the hash map is not tightly filled.
Therefore a hash map is appropriate if speed for access and insertions is important and
sufficient memory is available.
\begin{figure}
\begin{center}
\begin{tabular}{|lllll|}
\hline
 & & subscription & list operations & back operations \\
\hline
vector & one-dimensional array & $O(1)$ & $O(n)$ & $O(1)$ \\
list   & double-linked list    & $O(n)$ & $O(1)$ & $O(1)$ \\
map    & binary tree           & $O( \log(n) )$ & $O( \log(n) )$ & --- \\
map    & hash map              & $O( 1 )$ & $O( 1 )$ & --- \\
\hline
\end{tabular}
\caption{\label{containers} \it The cost for specific operations with different
types of containers. A dash indicates that the corresponding operation
is usually not provided for this container. For the hash map it is assumed that the map
is not tightly filled, such that collisions occur rarely.}
\end{center}
\end{figure}
\\
Fig. (\ref{containers}) summarizes the cost for specific operations
involving the various types of containers.
Back operations are operations, where access, insertion or deletion
occur at the end of the container. A typical example are stack operations.

\subsection{Object-oriented design: A little bit C++}

Object-oriented programming techniques were invented to allow the 
development and maintenance of large, complex software applications.
Experience showed that techniques which worked for smaller projects
do not necessarily scale to larger projects.
In this section I discuss some features of C++ as an example
for an object-oriented programming language.
A detailed introduction to C++ and object-oriented techniques
can be found in the book by Stroustrup \cite{Stroustrup}.
Most of the topics discussed here are also present in other 
object-oriented languages or can be modelled by the user
in languages which do not have native support for object-oriented techniques.
\\
There are several reasons, which motivate
the particular choice of C++ from the set of object-oriented programming languages:
Since 1998 C++ is standardized, which enhances the portability of programs to different
platforms. This is of particular importance in academia, where one moves in the early
stages of a career from one post-doc position to another.
Furthermore, C++ is widely used. As a consequence, additional development tools
are available. The list includes debuggers, editors with automatic high-lighting
facilities for C++ code and tools for the automatic generation of documentation.
Finally, C++ allows operator overloading. In the scientific domain, operations like
addition and multiplication are frequently used.
Operator overloading allows one to use the same notation for different data types,
e.g.  one can write for example 
\begin{verbatim}
           2*a + b
\end{verbatim}
independently if \v/a/ and \v/b/ are numbers, vectors or matrices.
This makes programs more readable.
It is also a major argument against Java in scientific computing.
Java does not allow operator overloading and writing
\begin{verbatim}
           v.add_vec(w);
\end{verbatim}
for the addition of two vectors makes programs less readable.
\\
A first tool for object-oriented programming is a modular approach
corresponding to a ``divide and conquer'' strategy.
If a problem can be cleary separated into two sub-problems, the complexity
is already greatly reduced.
A basic module or the smallest independent entity in C++ is a class.
A class consists of data members and methods 
operating on the data (methods are also called member functions).
An example for a class would be the complex numbers, whose data
members are two double precision variables $x$ and $y$, representing
the real and imaginary part, respectively.
Member functions could be a print routine, addition and subtraction, etc..
Within the ``divide and conquer'' strategy falls also the 
strict separation of the implementation from the interface.
This is done by private and public members. Data members are usually taken
to be private members, to protect them to be changed accidently from
outside the class.
A well designed separation between an interface and an implementation allows
one to replace the implementation of a specific method
by a more efficient algorithm without changing the interface.
Therefore no changes have to be done to the rest of the program.
\\
A second tool for object-oriented programming is data abstraction.
General purpose programming languages come usually with a limited
number of build-in data types (integers, real double precision, etc.).
In many cases the user would like to have additional data types, 
in high energy physics one might like to have a data type ``complex number''
or even a data type ``Feynman diagram''.
Now, there might be programming languages which have a build-in data type
``complex number'', but it is quite unlikely to find a language which
has a data type ``Feynman diagram''. 
The ability to extend the build-in data types by user-defined data types
is therefore essential.
Furthermore it occurs quite often that several data types are related to each
other.
In C++ similar concepts are modeled through inheritance and derived classes.
For example, if there are the data types ``strongly interacting particle'',
``gluon'' and ``quark'', the type ``strongly interacting particle'' would
be a base class from which the classes ``gluon'' and ``quark'' are derived.
Derived classes are specializations of base classes and may have additional
properties. Therefore, a derived class may be used where-ever only the base class
is required. 
Inheritance and derived classes can be used to avoid that similar code
is implemented more than once.
Within the C++ community one draws in inheritance diagrams arrows 
from the derived class pointing to the base class.
\\
A third tool of object-oriented programming is the ability to re-use
and extend a given program. 
The traditional re-use of an existing program consists in an application,
where the new code calls the old code.
Object-oriented programming techniques allow also the reverse situation, 
e.g. that the old code calls the new code.
In C++ this can be done through virtual functions which are resolved
at run-time.
A typical example is the situation where the old program is a larger framework,
which one would like to extend with some new functionality.
If implemented properly, this will work even without recompiling the old
code.
\\
The fourth tool of object-oriented programming techniques are generic
algorithms.
For example, sorting a list of integers or a list of real double precision
variables can be done with the same algorithm.
In C++ the use of templates allows the implementation of generic algorithms.
The Standard Template Library (STL) already contains implementations
for the most important algorithms, among other things it provides
vectors, lists, maps and hash maps, discussed in the previous section.
\\
C++ was developed from the C programming language and retains C as
a subset.
As in C, data can be stored in memory either automatically on the stack,
dynamically allocated on the heap
or statically at a fixed address.
For subroutine calls
it is usually not efficient to copy a large data structure to local
variables upon entering the subroutine.
Instead it is more appropriate to pass a reference or a pointer to 
the subroutine.

\subsection{A simple toy computer algebra system}

In this section I will discuss the design of a simple toy computer algebra
system.
This toy model should know about symbols (``a'', ``b'', etc.), integers
(``1'', ``2'', etc.), addition and multiplication.
It should be possible to enter an expression like
\bq
5 a + 3 b + 2 a
\eq
which is automatically simplified to $7 a + 3 b$. 
The discussion of the implementation for this toy program
follows closely the structure of the GiNaC-library 
\cite{Bauer:2000cp,Bauer:diplom}.
For the GiNaC-library the source code is freely available 
and the interested reader is invited
to study the techniques explained below in a real case example.
\\
From the description above we can identify five objects: symbols, integers,
addition, multiplication and expressions.
Symbols and integers cannot have any subexpressions and are atoms, 
addition and multiplication are containers.
An expression can be an atom or a container.
The natural design is therefore to start from a base class ``\v/basic/'',
representing a basic expression, from which the classes ``\v/symbol/'',
``\v/numeric/'', ``\v/add/'' and ``\v/mul/'' are derived.
The lay-out of the atomic classes is straightforward:
The class ``\v/symbol/'' must store it's name (a string like ``a'' or ``x''), whereas
the class ``\v/numeric/'' must store it's numeric value.
Since both are derived from the class ``\v/basic/'', the relevant part of code in C++
could look as follows:
\begin{verbatim}
class symbol : public basic
{
 protected:
        std::string name; 
};

class numeric : public basic
{
 protected:
        int value;
};
\end{verbatim}
Before dealing with the lay-out of the remaining classes (in particular with
the lay-out of the class ``\v/basic/''), it is worth
to consider memory management and efficiency first.
Already in our simple toy example, expressions can be nested and can become
large.
It is inefficient to copy large expressions upon entering or leaving
a subroutine.
Passing a pointer is more efficient.
However, programming with pointers at a high level is not elegant and
error-prone. One therefore hides all pointer operations into an
encapsulating class ``\v/ex/'':
\begin{verbatim}
class ex
{
 public:
        basic *bp;
};
\end{verbatim}
The user deals only with the class ``\v/ex/'' and never manipulates pointers
directly.
An instance of the class ``\v/ex/'' consists only of a pointer to 
a ``\v/basic/''-object and is therefore extremely light-weight.
When a new expression is created in a subroutine, this subroutine
returns an ``\v/ex/'', e.g. a pointer to some ``\v/basic/''-object.
This requires that the ``\v/basic/''-object is created dynamically on the heap.
If it would be created on the stack, it will get automatically destroyed
at the end of the subroutine, and the program will abort as soon as it
tries to access this object outside the subroutine.
However with dynamic memory allocation one has to ensure that objects,
which are no longer needed, get deleted and that the occupied memory space
is freed.
Otherwise one would run out of memory soon.
This can be done with a technique called ``reference counting with 
copy-on-write semantics'' \cite{Cline}.
The class ``\v/basic/'' has a counter, which keeps track of how many times
this ``\v/basic/''-object is pointed to:
\begin{verbatim}
class basic
{
 private:
        unsigned refcount;
};
\end{verbatim}
It happens quite often that some expression is assigned to more than one variable.
With reference counting, there is no need to store the same ``\v/basic/''-object
twice in memory.
In the following lines of code
\begin{verbatim}
   ex e1 = 3*a + 2*b;
   ex e2 = e1;
\end{verbatim}
both \v/e1/ and \v/e2/ point to the same ``\v/basic/''-object
and no copying takes place in the second line of this code.
After these lines have been executed, the 
\v/refcount/ variable of the ``\v/basic/''-object is equal to $2$.
However, copying is necesarry when one expression is changed:
\begin{verbatim}
   e2 = e2 + 4;
\end{verbatim}
This is called copy-on-write semantics.
In this example \v/e2/ pointed first to the expression $3a+2b$, and then
to the new expression $3a+2b+4$. The implementation ensures that the 
reference counter of $3a+2b$ is decreased by one, as soon as \v/e2/ points
to the new expression $3a+2b+4$.
Obviously, if the \v/refcount/ variable of some ``\v/basic/''-object equals
zero, it is no longer referenced by any variable and therefore it can 
be safely deleted. 
This releases the memory occupied by the ``\v/basic/''-object and avoids memory
leaks.
\\
\\
Let us now discuss the lay-out of the container classes ``\v/add/'' and ``\v/mul/''.
As an example we consider again the expression $3a+2b+4$. 
The naive representation would be a nested structure, where the top-level
container is of type ``\v/add/'' with three summands.
One summand is the numerical value $4$, while the other two summands are
of type ``\v/mul/'' and given by $3a$ and $2b$, respectively.
Unfortunately this representation will result in rather deep trees, which
are slow to manipulate.
Since products of the form ``numerical coefficient'' times ``something else''
occur quite often, it is advantageous to introduce an additional data
type for this pair:
\begin{verbatim}
class expair
{
 public:
        ex rest;    // first member of pair, an arbitrary expression
        ex coeff;   // second member of pair, must be numeric
};
\end{verbatim}
These pairs are easy to manipulate, for example, if two pairs in a sum have the
same value for the variable \v/rest/, one can simply add their numerical
coefficients.
The introduction of these pairs flattens expression trees significantly.
Our sum $3a+2b+4$ can now be represented by a sequence of two pairs 
$(a,3)$, $(b,2)$ and
the overall numerical constant $4$.
If one would introduce powers into the computer algebra system, a similar
structure is seen for products, for example $5 x^2 y^3 z$ can be
represented by the sequence of pairs $(x,2)$, $(y,3)$ and $(z,1)$
and the overall numerical constant $5$.
Again, if two pairs in the product
have the
same value for the variable \v/rest/, one can simply add the numerical
values of their variables \v/coeff/, which now represent the exponents.
It is therefore appropriate to introduce for the containers ``\v/add/'' and ``\v/mul/''
a common base class ``\v/expairseq/'', which implements this data structure, e.g.
a sequence of pairs together with an additional numerical coefficient:
\begin{verbatim}
class expairseq : public basic
{
 protected:
        std::vector<expair> seq;
        ex overall_coeff;
};
\end{verbatim}
The sequence of pairs is implemented by using a \v/vector/ from the
Standard Template Library.
The classes ``\v/add/'' and ``\v/mul/'' are then derived from ``\v/expairseq/'':
\begin{verbatim}
class add : public expairseq
{ };

class mul : public expairseq
{ };
\end{verbatim}
Note that ``\v/add/'' and ``\v/mul/'' do not need any additional data members.
Finally, we overload the operators $+$ and $*$:
\begin{verbatim}
const ex operator+(const ex & lh, const ex & rh);
const ex operator*(const ex & lh, const ex & rh);
\end{verbatim}
These operators call the constructors of the classes ``\v/add/'' and
``\v/mul/'', respectively.
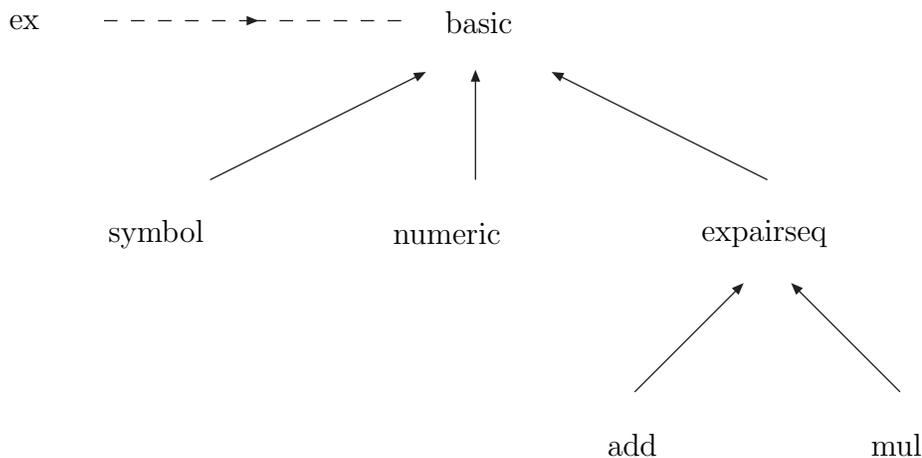
\begin{figure}
\begin{center}
\begin{picture}(400,180)(0,0)
\Text(20,170)[c]{ex}
\DashArrowLine(50,170)(162,170){5}
\Text(192,170)[c]{basic}
\Text(70,90)[c]{symbol}
\Text(180,90)[c]{numeric}
\Text(300,90)[c]{expairseq}
\Text(250,10)[c]{add}
\Text(350,10)[c]{mul}
\LongArrow(90,110)(170,150)
\LongArrow(190,110)(190,150)
\LongArrow(300,110)(220,150)
\LongArrow(250,30)(290,70)
\LongArrow(350,30)(310,70)
\end{picture}
\caption{\label{casclasses} \it The class structure of a simple toy computer
algebra system. Long arrows point from derived classes to the base classes.
A dashed arrow indicates that this class is a ``smart pointer''
to another class.}
\end{center}
\end{figure}
Fig. (\ref{casclasses}) summarizes the class hierarchy of the simple
toy computer algebra program.
Note that the class ``\v/expair/'' is an internal helper class and not
derived from ``\v/basic/''.
\\
\\
We now discuss how to implement automatic simplifications.
For example, we would like to have that $5 a + 3 b + 2 a$ 
is automatically evaluated to $7 a + 3 b$. 
To this aim we introduce for the class ``\v/basic/'' a status flag ``evaluated''
and a method \v/eval()/:
\begin{verbatim}
class basic
{
 public:
        virtual ex eval(void) const;
 protected:
        mutable unsigned flags;
};
\end{verbatim}
Every time a ``\v/basic/''-object is assigned, this flag is checked.
If the flag is not set, the method \v/eval()/ is called.
The base class ``\v/basic/'' has a trivial evaluation routine, which does
nothing except setting the flag to evaluated.
This implementation is also sufficient for the atomic classes
``\v/symbol/'' and ``\v/numeric/''.
The method \v/eval()/ is a virtual function and can be redefined
in derived classes.
Since all expressions are always accessed through pointers via the class
\v/ex/ the mechanism of virtual functions ensures that
the appropriate method is called.
In our example, the classes ``\v/add/'' and ``\v/mul/'' would redefine
the virtual function \v/eval/:
\begin{verbatim}
class add : public expairseq
{
 public:
        ex eval(void) const;
}
\end{verbatim}
An implementation of the method \v/eval()/ for the class ``\v/add/'' would
sort the elements of the sequence of ``\v/expair/''s and combine element which
have the same ``\v/rest/''.
Note that automatic evaluation happens the first time an object is assigned,
not the first time an object is constructed.
This is due to the fact, that an object of type ``\v/add/'' like $2x-x$ 
evaluates to $x$, which is of type ``\v/symbol/'' and not of type ``\v/add/''.

\section{Efficiency}

This section is devoted to issues related to efficiency.
I discuss three examples: Shuffling relations, fast multiplication
and the calculation of the greatest common divisor.
Shuffling relations are discussed as an application of recursive
techniques.
For the multiplication of large numbers I outline three methods:
the classical text-book method, Karatsuba's algorithm and a method
based on fast Fourier transform.
An algorithm for the calculation for the greatest common denominator
is of central importance for many other algorithms.
Therefore the Euclidean algorithm and improvements are discussed in detail.

\subsection{Shuffling}
\label{sec:shuffling}

Recursive techniques are often used in symbolic calculations.
A classical example are
the Fibonacci numbers defined by $f(0)=1$, $f(1)=1$ and
\bq
\label{fibonacci}
f(n) & = & f(n-1) + f(n-2)
\eq
for $n \ge 2$.
Recursive procedures are easily implemented, but it should be noted
that in most cases they provide not an efficient way to solve a problem.
For example, the recursive definition of the Fibonacci numbers is 
certainly the most efficient way to create a table with the first $n$
Fibonacci numbers, but a recursive approach is highly inefficient, if
we just need one Fibonacci number $f(n)$, if $n$ is large.
For the Fibonacci numbers a closed formula is known,
\bq
f(n) & = & \frac{1}{\sqrt{5}} \left[ \left( \frac{1+\sqrt{5}}{2} \right)^{n+1}
                                    -\left( \frac{1-\sqrt{5}}{2} \right)^{n+1}
                              \right],
\eq
and it is more efficient to use this closed formula in the latter case.
\\
With this warning, I will now discuss some issues related to the implementation
of recursive procedures in computer algebra systems.
As an example I will use shuffling relations.
Shuffling relations occur frequently in calculations.
I discuss them for Euler-Zagier sums, which are defined by
\bq 
\label{EulerZagier}
  Z_{m_1,...,m_k}(n) & = & \sum\limits_{n\ge i_1>i_2>\ldots>i_k>0}
     \frac{1}{{i_1}^{m_1}}\ldots \frac{1}{{i_k}^{m_k}}.
\eq
Euler-Zagier sums form an algebra, that is to say that the product of two
Euler-Zagier sums with the same upper summation limit $n$ 
is again a finite sum of single Euler-Zagier sums.
The product can be worked out recursively via the formula
\bq
\label{Zmultiplication}
\lefteqn{
Z_{m_1,...,m_k}(n) \times Z_{m_1',...,m_l'}(n) } & & \nonumber \\
& = & \sum\limits_{i_1=1}^n \frac{1}{i_1^{m_1}} Z_{m_2,...,m_k}(i_1-1) Z_{m_1',...,m_l'}(i_1-1) \nonumber \\
&  & + \sum\limits_{i_2=1}^n \frac{1}{i_2^{m_1'}} Z_{m_1,...,m_k}(i_2-1) Z_{m_2',...,m_l'}(i_2-1) \nonumber \\
&  & + \sum\limits_{i=1}^n \frac{1}{i^{m_1+m_1'}} Z_{m_2,...,m_k}(i-1) Z_{m_2',...,m_l'}(i-1).
\eq
This multiplication can be implemented by a subroutine, which takes three lists
\v/res/, \v/arg1/ and \v/arg2/
as arguments.
When the routine is entered the first time, \v/res/ is empty and \v/arg1/ and \v/arg2/ contain
$(m_1,...,m_k)$ and $(m_1',...,m_l')$, respectively.
If \v/arg1/ (or \v/arg2/) is the empty list, we are basically done: 
We append the content of \v/arg2/ (or of \v/arg1/)
to \v/res/ and return \v/res/.
Otherwise we remove the first element from \v/arg1/ and append it to \v/res/ and use recursion,
this corresponds to the first line on the r.h.s of eq. (\ref{Zmultiplication}).
Similar, for the second (or third line) of eq. (\ref{Zmultiplication}) one removes the first
element from \v/arg2/, (or the first elements from \v/arg1/ and \v/arg2/) and appends
it (or the sum of the two elements) to \v/res/.
Most computer algebra systems provide a list data type within their framework.
In GiNaC this type is called \v/lst/ and is a derived class from \v/basic/.
It is tempting to implement the shuffle multiplication using this data type.
In GiNaC this could look as follows:
\begin{verbatim}
ex shuffle_mul(const ex & res, const ex & arg1, const ex & arg2);
\end{verbatim}
where it is understood that \v/res/, \v/arg1/ and \v/arg2/ will always point to a \v/lst/.
However, such an approach can lead to a considerable performance loss, since the arguments
\v/res/, \v/arg1/ and \v/arg2/ are now under the spell of the automatic evaluation procedure.
The algorithm according to eq. (\ref{Zmultiplication}) shuffles basically elements from \v/arg1/
and \v/arg2/ to \v/res/. At each step new lists are created with appended or removed elements,
and the automatic evaluation procedure will check for each new list, if all of its elements
are already evaluated. Usually all elements are already evaluated, therefore this procedure
is a waste of computer resources.
It is therefore better to use for the lists a temporary data structure 
which is not related to
the automatic evaluation procedure.
In our case, a vector containing the type \v/ex/ would do the job:
\begin{verbatim}
typedef std::vector<ex> exvector;

ex shuffle_mul(const exvector & res, const exvector & arg1, 
                                     const exvector & arg2);
\end{verbatim}
Note that in the discussion of the GiNaC-library we already crossed the data structure
\v/expair/, which is not derived from the class \v/basic/ and therefore not subject
to the automatic evaluation procedure.
\\
Situations where this technique can be used occur quite frequently.
Another example is the shuffle product of iterated integrals.
Let us consider the integral
\bq
\label{Gfunc}
G(z_1,...,z_k;y) & = &
 \int\limits_0^y \frac{dt_1}{t_1-z_1}
 \int\limits_0^{t_1} \frac{dt_2}{t_2-z_2} ...
 \int\limits_0^{t_{k-1}} \frac{dt_k}{t_k-z_k}.
\eq
A product of two such integrals can be reduced to a sum of single integrals
according to
\bq
G(z_1,...,z_k;y) \times G(z_{k+1},...,z_n;y)
 & = & \sum\limits_{shuffles} G(z_{\sigma(1)},...,z_{\sigma(n)},y),
\eq
where the sum is over all permutations of $z_1$, ..., $z_n$, which keep
the relative order of $z_1$, ..., $z_k$ and $z_{k+1}$, ..., $z_n$ fixed.
Recursively, we can write for the product
\bq
\lefteqn{
G(z_1,...,z_k;y) \times G(z_{k+1},...,z_n;y) = } \\
 & & 
 \int\limits_0^y \frac{dt}{t-z_1} G(z_2,...,z_k;t) G(z_{k+1},...,z_n;t) 
 + \int\limits_0^y \frac{dt}{t-z_{k+1}} G(z_1,...,z_k;t) G(z_{k+2},...,z_n;t)
 \nonumber
\eq
and the algorithm can be implemented in complete analogy to the 
one for eq. (\ref{Zmultiplication}).
\\
Similar considerations apply if we just want to transform one representation to another
form.
An examples is a change of basis from Euler-Zagier sums to harmonic sums, defined by
\bq
S_{m_1,...,m_k}(n)  & = & 
\sum\limits_{n\ge i_1 \ge i_2\ge \ldots\ge i_k \ge 1}
\frac{1}{{i_1}^{m_1}}\ldots \frac{1}{{i_k}^{m_k}}.
\eq
The difference between Euler-Zagier sums and harmoinc sums is the upper summation
limit for the subsums: $i-1$ for Euler-Zagier sums and $i$ for harmonic sums.
The conversion between the two bases uses the formulae
\bq
\label{conversion}
Z_{m_1,...,m_k}(n) & = & 
 \sum\limits_{i_1=1}^n \frac{1}{{i_1}^{m_1}} Z_{m_2,...,m_k}(i_1) 
 \;\; - \;\; Z_{m_1+m_2,m_3,...,m_k}(n),
\nonumber \\ 
S_{m_1,...,m_k}(n) & = & 
 \sum\limits_{i_1=1}^n \frac{1}{{i_1}^{m_1}} S_{m_2,...,m_k}(i_1-1) 
 \;\; + \;\; S_{m_1+m_2,m_3,...,m_k}(n).
\eq
and are implemented by routines taking two arguments:
\begin{verbatim}
ex Zsum_to_Ssum(const exvector & res, const exvector & arg);
ex Ssum_to_Zsum(const exvector & res, const exvector & arg);
\end{verbatim}

\subsection{Multiplication of large numbers}

In this paragraph I discuss issues related to the efficiency for the multiplication
of large numbers.
Multiprecision arithmetic is needed in many applications.
For the practitioner, there are several freely-available libraries which provide
support for multiprecision numbers. Examples are the libraries
GMP \cite{GMP}, CLN \cite{CLN} or NTL \cite{NTL}.
\\
\\
Let $a$ be an integer with $n$ digits in the base $B$, e.g.
\bq
a & = & a_0 + a_1 B + ... + a_{n-1} B^{n-1}.
\eq
For simplicity we assume that $n$ is even (if $n$ is odd we can just add a zero to the front).
We can write $a$ as
\bq
a & = & a_h B^{n/2} + a_l,
\eq
where $a_h$ and $a_l$ have now $n/2$ digits in base $B$.
Let $b$ be another integer with $n$ digits. The product $a b$ can be calculated as
\bq
\label{classmult}
 a b & = & a_h b_h B^n + \left( a_h b_l + a_l b_h \right) B^{n/2} + a_l b_l
\eq
This is the classical method. To multiply two integers with $n$ digits requires 4 multiplications
of integers with $n/2$ digits and four additions.
Usually the computational cost of the additions can be neglected against the one for the multiplications.
It is not too hard to see that the classical method is of order $O(n^2)$.\\
\\
The efficiency can be improved for large integers by rewritting eq. (\ref{classmult}) as follows:
\bq
\label{karatsuba}
 a b & = & a_h b_h B^n + \left[ a_h b_h + a_l b_l - \left(a_h - a_l \right) \left(b_h - b_l \right) \right] B^{n/2} + a_l b_l
\eq
This method requires only 3 multiplications of integers with $n/2$ digits and was invented by Karatsuba \cite{Karatsuba}.
It can be shown that this algorithms grows like $O(n^{\log_2 3}) \approx O(n^{1.58})$ with the number of digits $n$.
\\
\\
There is even a faster algorithm developed by Sch\"onhage and Strassen \cite{Schoenhage}
and based
on the fast Fourier transform.
It is most easily explained for the multiplication of polynomials.
A polynomial of degree $n$ is usually represented by its coefficients $a_i$:
\bq
a(x) & = & a_0 + a_1 x + a_2 x^2 + ... + a_n x^n.
\eq
However, a polynomial of degree $n$ is also uniquely defined by the values
at $n+1$ distinct points $x_0$, $x_1$, ..., $x_n$.
This representation is called the modular representation.
Suppose that we would like to multiply a polynomial $a(x)$ of degree $n$
by a polynomial $b(x)$ of degree $m$.
Then the product is of degree $n+m$.
Suppose that we know the polynomials $a(x)$ and $b(x)$ 
at $n + m +1$ distinct points:
\bq
a(x) & = & \left( a(x_0), ..., a(x_{n + m}) \right), \nonumber \\
b(x) & = & \left( b(x_0), ..., b(x_{n + m}) \right).
\eq
Then 
\bq
\left( \; a(x_0) \cdot b(x_0), \; ..., \; a(x_{n + m}) \cdot b(x_{n + m}) \; \right)
\eq
defines uniquely the polynomial $a(x) b(x)$ and requires only
$O(n+m)$ operations.
In the modular representation, multiplication of two polynomials of degree $n$ is an
$O(n)$ operation.
However, this is only the cost if the objects are already in this particular
representation.
In general, we have to add the costs for converting the input polynomials to
the modular representations and the cost for converting the result back
to a standard representation.
The method will be  competitive only if these transformations can be done
at a cost lower than, say, Karatsuba multiplication.
Using a method based on the fast Fourier transform, it turns out that these 
transformations can be done efficiently and that the cost for these transformations
is $O(n \log n )$.
Let us first consider the transformation from the standard representation
\bq
a(x) & = & a_0 + a_1 x + ... + a_{n-1} x^{n-1}
\eq
of a polynomial of degree less than $n$ to the modular represenation.
For simplicity we assume that $n$ is even.
We have the freedom to chose the $n$ distinct evaluation points and a clever choice
can reduce the required amount of calculation.
A suitable choice is to evaluate the polynomial
at the special points
$\left\{ 1, \omega , \omega^2 ,..., \omega^{n-1} \right\}$,
where $\omega$ is a primitive $n$-th root of unity, e.g.
\bq
\omega^n = 1, & & \mbox{but} \; \omega^k \neq 1 \; \mbox{for} \; 0 < k < n.
\eq
A few examples for primitive roots of unity:
In ${\mathbb C}$, $i$ and $-i$ are primitive $4$-th roots of unity, $-1$ is not
a primitive $4$-th root of unity, since $(-1)^2=1$.
In ${\mathbb Z}_{17}$, $4$ is a $4$-th root of unity, since 
$4^4 = 1 \; \mbox{mod} \;17$ and
$4^2 = 16 \; \mbox{mod} \;17$ as well as
$4^3 = 13 \; \mbox{mod} \;17$.
\\
The evaluation of the polynomial at this particular set of evaluation points is nothing
than performing a discrete Fourier transform.
Given a finite set of values $(a_0, a_1, ..., a_{n-1})$, the discrete Fourier transform
maps those values to the set $(\hat{a}_0, \hat{a}_1, ..., \hat{a}_{n-1} )$, where
\bq
\hat{a}_i & = & \sum\limits_{j=0}^{n-1} a_j \omega^{i j}.
\eq
An efficient way to calculate the values $(\hat{a}_0, \hat{a}_1, ..., \hat{a}_{n-1} )$
is given by the fast Fourier transform.
Since we assumed that $n$ is even, the $n$-th roots of unity satisfy
\bq
\omega^{i+\frac{n}{2}} & = & - \omega^i.
\eq
We write the polynomial $a(x)$ in the form
\bq
\label{FFT}
a(x) & = & b(x^2) + x \cdot c(x^2),
\eq
where the polynomials $b(y)$ and $c(y)$ have at most one-half the degree of $a(x)$.
Since we have $\left( \omega^{i+n/2}\right)^2 = \left( \omega^i \right)^2$
it is sufficient to evaluate the polynomials $b(y)$ and $c(y)$ at $n/2$ distinct points
instead of $n$.
This is the basic building block for the fast Fourier transform.
Using this technique recursively yields a method which performs the Fourier transform
at $O(n \log n)$.
In complete analogy, the conversion from the modular representation to the
standard representation is done by the inverse discrete Fourier transform.
Given a finite set of values $(\hat{a}_0, \hat{a}_1, ..., \hat{a}_{n-1} )$, the inverse 
discrete Fourier transform
maps those values to the set $(a_0, a_1, ..., a_{n-1})$, where
\bq
a_j & = & \frac{1}{n} \sum\limits_{k=0}^{n-1} \hat{a}_k \omega^{- j k}.
\eq
Again, a similar decomposition like in eq. (\ref{FFT}) is used to speed up the calculation.

\subsection{The greatest common divisor and the Euclidean algorithm}

It is often required to simplify rational functions by canceling common factors in the numerator
and denominator.
As an example let us consider
\bq
\frac{(x+y)^2 (x-y)^3}{(x+y)(x^2-y^2)} 
 & = & (x-y)^2.
\eq
One factor of $(x+y)$ is trivially removed, the remaining factors are cancelled
once we noticed that $(x^2-y^2) = (x+y)(x-y)$.
For the implementation in a computer algebra system this is however not the way to proceed.
The factorization of the numerator and the denominator into irreducible polynomials
is a very expensive calculation and actually not required.
To cancel the common factors in the numerator and in the denominator it is sufficient to 
calculate the greatest common divisor (gcd) of the two expressions.
The efficient implementation of an algorithm for the calculation of the gcd is essential
for many other algorithms.
Like in the example above, most gcd calculations are done in polynomial rings.
It is therefore useful to recall first some basic definitions from ring theory:
\\
A commutative ring $(R,+,\cdot)$ is a set $R$ with two operations $+$ and $\cdot$, such 
that $(R,+)$ is an abelian group and $\cdot$ is associative, distributive and commutative.
In addition we always assume that there is a unit element for the multiplication.
An example for a commutative ring would be ${\mathbb Z}_8$, e.g. the set 
of integers modulo 8.
In this ring one has for example $3+7 = 2$ and $2 \cdot 4 = 0$.
From the last equation one sees that it is possible to obtain zero
by multiplying two non-zero elements.
\\
An integral domain is a commutative ring with the additional requirement
\bq
a \cdot b = 0 & \Rightarrow & a=0 \;\; \mbox{or} \;\; b=0 \;\;\; \mbox{(no zero divisors)}.
\eq
Sometimes an integral domain $D$ is defined by requiring
\bq
a \cdot b = a \cdot c \;\;\mbox{and}\;\; a \neq 0 
 & \Rightarrow & b=c \;\;\; \mbox{(cancellation law)}.
\eq
It can be shown that these two requirements are equivalent.
\begin{figure}
\begin{center}
\begin{picture}(200,180)(0,0)
\Text(100,10)[c]{Field}
\Text(100,50)[c]{Euclidean domain}
\Text(100,90)[c]{Unique factorization domain}
\Text(100,130)[c]{Integral domain}
\Text(100,170)[c]{Commutative ring}
\LongArrow(100,20)(100,40)
\LongArrow(100,60)(100,80)
\LongArrow(100,100)(100,120)
\LongArrow(100,140)(100,160)
\end{picture}
\caption{\label{domains} \it Hierarchy of domains. Upward pointing arrows indicate
that a former domain is automatically also a latter domain.}
\end{center}
\end{figure}
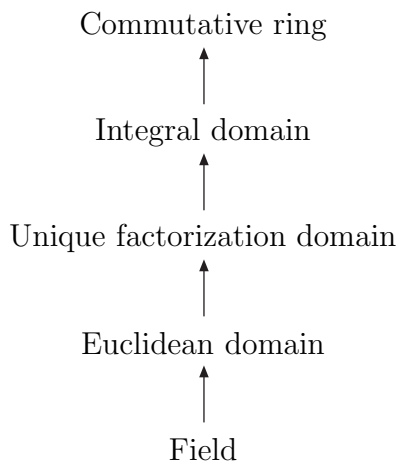
An example for an integral domain would be the subset of the complex numbers
defined by
\bq
\label{exampleintdomain}
S & = & \left\{ \left. a + b i \sqrt{5} \; \right| a,b, \in {\mathbb Z} \right\}
\eq
An element $u \in D$ is called unit or invertible if $u$ has a
multiplicative inverse in $D$.
The only units in the example eq. (\ref{exampleintdomain}) are $1$ and $-1$.
We further say that $a$ divides $b$ if there is an element $x \in D$ such that
$b = a x$. In that case one writes $a | b$.
Two elements $a,b \in D$ are called associates if $a$ divides $b$ and
$b$ divides $a$.
\\
We can now define the greatest common divisor (gcd): An element $c \in D$ is called
the greatest common divisor of $a$ and $b$ if $c | a$ and $c | b$ and $c$ is a multiple
of every other element which divides both $a$ and $b$.
Closely related to the gcd is the least common multiple (lcm) of two elements $a$ and $b$:
$d$ is called least common multiple of $a$ and $b$ if $a | d$ and $b | d$ and $d$ is a divisor
of every other element which is a multiple of both $a$ and $b$.
Since gcd and lcm are related by
\bq
\mbox{lcm}(a,b) & = & \frac{a b}{\mbox{gcd}(a,b)}
\eq
it is sufficient to focus on an algorithm for the calculation of the gcd. 
\\
\\
An element $p \in D-\{0\}$ is called prime if
from $p | a b$ it follows that $p | a$ or $p | b$.
An element $p \in D-\{0\}$ is called irreducible if
$p$ is not a unit and whenever
$p=a b$ either $a$ or $b$ is a unit.
In an integral domain, any prime element is automatically also an irreducible
element.
However, the reverse is in general not true.
This would require some additional properties in the ring.
\\
An integral domain $D$ is called a unique factorization domain if for
all $a\in D-\{0\}$, either $a$ is a unit or else $a$ can be expressed as a finite product
of irreducible elements
such that this factorization into irreducible elements is unique up to associates and reordering.
It can be shown that in an unique factorization domain the notions of irreducible element and
prime element are equivalent.
In a unique factorization domain the gcd exists and is unique (up to associates and reordering).
The integral domain $S$ in eq. (\ref{exampleintdomain}) is not a unique
factorization domain, since for example
\bq
 21 & = & 3 \cdot 7 = \left( 1 - 2 i \sqrt{5} \right) \left( 1 + 2 i \sqrt{5} \right)
\eq 
are two factorizations into irreducible elements. An example for a unique
factorization domains is the polynomial ring ${\mathbb Z}[x]$ in one variable with integer
coefficients.
\\
\\
An Euclidean domain is an integral domain $D$ 
with a valuation map $v: D-\{0\} \rightarrow {\mathbb N}$ into the nonnegative integer numbers, such that
$v(a b) \ge v(a)$ for all $a,b \in D -\{0\}$, and for all
$a,b \in D$ with $b \neq 0$, there exist elements $q,r \in D$ 
such that
\bq
a = bq +r,
\eq
where either $r=0$ or $v(r) < v(b)$.
This means that in an Euclidean domain division with remainder is possible.
An example for an Euclidean domain is given by the integer numbers ${\mathbb Z}$.
\\
\\
Finally, a field is a commutative ring in which every nonzero element has a multiplicative inverse,
e.g. $R-\{0\}$ is an abelian group.
Any field is automatically an Euclidean domain.
Examples for fields are given by the rational numbers ${\mathbb Q}$, 
the real numbers ${\mathbb R}$, the complex numbers ${\mathbb C}$
or ${\mathbb Z}_p$, the integers modulo $p$ with $p$ a prime number.
\begin{figure}
\begin{center}
\begin{tabular}{l|l|l}
 $R$ & $R[x]$ & $R[x_1,x_2,...,x_n]$ \\
\hline
 & \\
 commutative ring & commutative ring & commutative ring \\
 integral domain  & integral domain  & integral domain  \\
 unique factorization domain & unique factorization domain & unique factorization domain \\
 euclidean domain & unique factorization domain & unique factorization domain \\
 field            & euclidean domain & unique factorization domain \\
\end{tabular}
\caption{\label{unipolring} \it Structure of polynomial rings in one variable and several variables
depending on the
underlying coefficient ring $R$.}
\end{center}
\end{figure}
\\
Fig. (\ref{domains}) summarizes the relationships between the various domains.
Of particular importance are polynomial rings in one or several variables.
Fig. (\ref{unipolring}) summarizes the structure of these
domains.
Note that a multivariate polynomial ring $R[x_1,...,x_n]$ can always be viewed
as an univariate polynomial ring in one variable $x_n$ with coefficients in the
ring $R[x_1,...,x_{n-1}]$.
\\
\\
The algorithm for the calculation of the gcd in an Euclidean domain dates back
to Euclid \cite{Euclid}. 
It is based on the fact that if $a = b q + r$, then
\bq
\mbox{gcd}(a,b) & = & \mbox{gcd}(b,r).
\eq
This is easily seen as follows: Let $c=\mbox{gcd}(a,b)$ and $d=\mbox{gcd}(b,r)$.
Since $r=a-b q$ we see that $c$ divides $r$, therefore it also divides $d$.
On the other hand $d$ divides $a=b q + r$ and therefore it also divides $c$.
We now have $c|d$ and $d|c$ and therefore $c$ and $d$ are associates.
\\
It is clear that for $r=0$, e.g. $a = b q$ we have $\mbox{gcd}(a,b)=b$.
Let us denote the remainder as $r=\mbox{rem}(a,b)$.
We can now define a sequence
$r_0 = a$, $r_1 = b$ and $r_i=\mbox{rem}(r_{i-2},r_{i-1})$ for $i \ge 2$.
Then there is a finite index $k$ sucht that $r_{k+1}=0$ 
(since the valuation map applied to the remainders is a strictly decreasing function).
We have
\bq
\mbox{gcd}(a,b) & = & \mbox{gcd}(r_0,r_1) = \mbox{gcd}(r_1,r_2)
 = ... = \mbox{gcd}(r_{k-1},r_k) = r_k.
\eq
This is the Euclidean algorithm.
We briefly mention that as a side product one can find elements $s$, $t$ such that
\bq
s a + t b & = & \mbox{gcd}(a,b).
\eq
This allows the solution of the Diophantine equation
\bq
s a + t b & = & c,
\eq
for $s$ and $t$ whenever $\mbox{gcd}(a,b)$ divides $c$.
\\
We are primarily interested in gcd computations in polynomial rings.
However, polynomial rings are usually only unique factorization domains,
but not Euclidean domains, e.g. division with remainder is in general not possible.
As an example consider the polynomials $a(x) = x^2+2x+3$ and $b(x)=5x+7$ in ${\mathbb Z}[x]$. 
It is not possible to write $a(x)$ in the form
$a(x) = b(x) q(x) + r(x)$, where the polynomials $q(x)$ and $r(x)$ have
integer coefficients.
However in ${\mathbb Q}[x]$ we have
\bq
x^2+2x+3 & = & \left( 5 x + 7 \right) \left( \frac{1}{5} x + \frac{3}{25} \right)
 + \frac{54}{25}
\eq
and we see that the obstruction arrises from the leading coefficent of $b(x)$.
It is therefore appropriate to introduce a pseudo-division with remainder.
Let $D[x]$ be a polynomial ring over a unique factorization doamin $D$.
For $ a(x) = a_n x^n + ... + a_0$, $b(x) = b_m x^m + ... + b_0$ with $n \ge m$ 
and $b(x) \neq 0$ there exists $q(x), r(x) \in D[x]$ such that
\bq
b_m^{n-m+1} a(x) & = & b(x) q(x) + r(x)
\eq
with $\mbox{deg}(r(x)) < \mbox{deg}(b(x))$.
This pseudo-divison property is sufficient to extend the Euclidean algorithm to
polynomial rings over unique factorization domains.
\\
\\
Unfortunately, the Euclidean algorithm as well as the extended algorithm with pseudo-division
have a severe drawback: Intermediate expressions can become quite long.
This can be seen in the following example, where we would like to calculate the 
gcd of the polynomials
\bq
a(x) & = & x^8 + x^6 - 3 x^4 - 3 x^3 + 8 x^2 + 2 x -5, \nonumber \\
b(x) & = & 3 x^6 + 5 x^4 - 4 x^2 - 9 x + 21,
\eq
in ${\mathbb Z}[x]$.
Calculating the pseudo-remainder sequence $r_i(x)$ we obtain
\bq
r_2(x) & = & -15x^4 + 3x^2 -9, \nonumber \\
r_3(x) & = & 15795 x^2 + 30375 x - 59535, \nonumber \\
r_4(x) & = & 1254542875143750 x - 1654608338437500, \nonumber \\
r_5(x) & = & 12593338795500743100931141992187500.
\eq
This implies that $a(x)$ and $b(x)$ are relatively prime, but the numbers which occur
in the calculation are large.
An analysis of the problem shows, that the large numbers can be avoided if each
polynomial is split into a content part and a primitive part.
The content of a polynomial is the gcd of all it's coefficients.
For example we have
\bq
15795 x^2 + 30375 x - 59535 & = & 1215 \left( 13 x^2 + 25 x + 49 \right)
\eq
and $1215$ is the content and $13 x^2 + 25 x + 49$ the primitive part.
Taking out the content of a polynomial in each step requires a gcd calculation in the
coefficient domain and avoids large intermediate expressions in the example above.
However the extra cost for the gcd calculation in the coefficient domain is prohibitive
for multivariate polynomials.
The art of gcd calculations consists in finding an algorithm which keeps intermediate
expressions at reasonable size and which at the same time does not involve too much
computational overhead.
An acceptable algorithm is given by the subresultant method \cite{Collins:srm,Brown:srm}:
Similar to the methods discussed above, one calculates a polynomial remainder 
sequence $r_0(x)$, $r_1(x)$, ... $r_k(x)$.
This sequence is obtained through
$r_0(x) = a(x)$, $r_1(x)=b(x)$ and
\bq
c_i^{\delta_i+1} r_{i-1}(x) & = & q_i(x) r_i(x) + d_i r_{i+1}(x)
\eq
where $c_i$ is the leading coefficient of $r_i(x)$, 
$\delta_i = \mbox{deg}(r_{i-1}(x)) - \mbox{deg}(r_{i}(x))$
and
$d_1 = (-1)^{\delta_1+1}$,
$d_i = - c_{i-1} \psi_i^{\delta_i}$ for $2 \le i \le k$.
The $\psi_i$ are defined by $\psi_1 = -1$ and
\bq
\psi_i & = & \left( - c_{i-1} \right)^{\delta_{i-1}} \psi_{i-1}^{1-\delta_{i-1}}.
\eq
Then the primitive part of the last non-vanishing remainder equals the primitive
part of $\mbox{gcd}(a(x),b(x))$.
\\
\\
I finally discuss an heuristic algorithm for the calculation of polynomial gcds \cite{Char}.
In general an heuristic algorithm maps a problem to a simpler problem, solves
the simpler problem and tries to reconstruct the solution of the original problem
from the solution of the simpler problem.
For the calculation of polynomial gcds one evaluates the polynomials at a specific point
and one considers the gcd of the results in the coefficient domain.
Since gcd calculations in the coefficient domain are cheaper, this can lead to a
sizeable speed-up, if both the evaluation of the polynomial and the reconstruction
of the polynomial gcd can be done at reasonable cost.
Let us consider the polynomials
\bq
a(x) & = & 6 x^4 + 21 x^3 + 35 x^2 + 27 x + 7, \nonumber \\
b(x) & = & 12 x^4 - 3 x^3 -17 x^2 - 45 x + 21.
\eq
Evaluating these polynomials at the point $\xi = 100$ yields
$a(100) = 621352707$ and $b(100)=1196825521$.
The gcd of theses two numbers is
\bq
c & = & \mbox{gcd}(621352707,1196825521) = 30607.
\eq
To reconstruct the polynomial gcd one writes $c$ in the
$\xi$-adic representation
\bq
c & = & c_0 + c_1 \xi + ... + c_n \xi^n, \;\;\; - \frac{\xi}{2} < c_i \le \frac{\xi}{2}.
\eq
Then the candidate for the polynomial gcd is
\bq
g(x) & = & c_0 + c_1 x + ... + c_n x^n.
\eq
In our example we have
\bq
30607 & = & 7 + 6 \cdot 100 + 3 \cdot 100^2
\eq
and the candidate for the polynomial gcd is $g(x) = 3 x^2 +6 x +7$.
A theorem guarantees now if $\xi$ is chosen such that
\bq
\xi > 1 + 2 \; \mbox{min}\left( ||a(x)||_\infty, ||b(x)||_\infty \right),
\eq
then $g(x)$ is the greatest common divisor of $a(x)$ and $b(x)$ if and only if
$g(x)$ divides $a(x)$ and $b(x)$.
This can easily be checked by a trial division.
In the example above, $g(x) = 3 x^2 +6 x +7$ divides both $a(x)$ and $b(x)$ and is therefore
the gcd of the two polynomials.
\\
Note that there is no guarantee that the heuristic algorithm will succeed in finding
the gcd.
But if it does, this algorithm is usually faster than the subresultant algorithm discussed
previously.
Therefore the strategy employed in computer algebra systems like Maple or GiNaC is to
try first a few times the heuristic algorithm with various evaluation points and to fall
back onto the subresultant algorithm, if the gcd has not been found by the 
heuristic algorithm.

\subsection{Remarks}

In this section we discussed issues related to efficiency.
\begin{figure}
\begin{center}
\begin{picture}(300,100)(0,0)
\Text(30,90)[c]{original problem}
\Text(30,10)[c]{simpler problem}
\Text(270,90)[c]{solution of original problem}
\Text(270,10)[c]{solution of simpler problem}
\ArrowLine(30,80)(30,20)
\ArrowLine(80,10)(180,10)
\ArrowLine(270,20)(270,80)
\Text(20,50)[r]{reduction}
\Text(280,50)[l]{reconstruction}
\Text(130,17)[b]{computation}
\end{picture}
\caption{\label{modularappr} \it The modular approach: Starting from the original problem,
one first tries to find a related simpler problem. The solution of the simpler problem
is used to reconstruct a solution of the original problem.}
\end{center}
\end{figure}
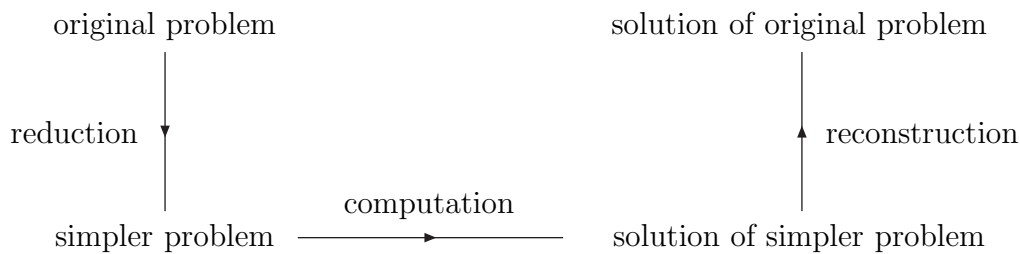
It is worth to recall two general strategies:
\begin{itemize}
\item The first one is the ``divide-and-conquer'' approach.
If a large task can be partitioned recursively into smaller units, this can lead to a considerable
speed-up.
We have seen examples for this approach in Karatsuba's algorithm, in the algorithm for
the fast Fourier transform and in the search for elements in a binary tree.
\item The second strategy is the modular technqiue.
Instead of solving a complicated problem right from the start, one tries
first to find a related simpler problem. 
From the solution of the simpler problem, one tries then
to reconstruct the solution of the original
problem.
Since the reduction to the simpler problem might involve some information loss, the reconstruction
of the full solution may not be possible or may not be unique.
However, if there is a simple and efficient way to check, if a guess for a solution is the correct
solution, this method can be highly competitive.
Fig. (\ref{modularappr}) summarizes this approach in a diagram.
We discussed as an example for this strategy the heuristic gcd algorithm.
\end{itemize}

\section{Classical Algorithms}
\label{sec:classical}

The development of computer algebra systems has triggered research in mathematics
on constructive algorithms for the solution of some important problems.
Examples are the factorization of polynomials, symbolic integration 
and symbolic summation, and simplifications with the help of Gr\"obner bases.
Most of these topics involve a big deal of mathematics.
Here I can only sketch an outline of the algorithms.

\subsection{Factorization}

It occurs frequently that we would like to factorize a polynomial.
Here I discuss the factorization of a polynomial $u(x) \in {\mathbb Z}[x]$
over the integers ${\mathbb Z}$.
I outline one specific algorithm due to Berlekamp.
The algorithm for the factorization is divided into three steps.
The first (and relative simple) step factors out the
greatest common divisior of the coefficients  and performs a
square-free decomposition.
The second step uses Berlekamp's algorithm and factors the polynomial
in the ring ${\mathbb Z}_p$, where $p$ is a prime number.
If $p$ were sufficiently large, such that all products of coefficients
would always lie between $-p/2$ and $p/2$, the factorization over
${\mathbb Z}$ could be read off from the factorization
over ${\mathbb Z}_p$.
Unfortunately, the required bound on $p$ can become rather large.
One uses therefore a small prime number $p$ and reconstructs in step 3
from a factorization over ${\mathbb Z}_p$ a factorization
over ${\mathbb Z}_{p^r}$, using a technique called Hensel lifting.
Since $r$ can be chosen large, this will yield a factorization over
${\mathbb Z}$.
\\
It should be noted that there are also other (and more efficient) algorithms
for the factorization of polynomials.
Examples are the probabilistic algorithm by Cantor and Zassenhaus \cite{Cantor}
or the algorithm by Kaltofen and Shoup \cite{Kaltofen}.
The computer program NTL by Shoup \cite{NTL}
provides a state-of-the-art implementation for the factorization
of univariate polynomials.

\subsubsection{Square-free decomposition}

Suppose a polynomial $u(x)$ contains a factor $v(x)$ to some power $m$,
e.g.
\bq
\label{squarefree1}
u(x) = \left( v(x) \right)^m r(x).
\eq
The exponent $m$ can easily be reduced to one as follows:
One starts by forming the derivative
\bq
\label{squarefree2}
u'(x) & = & m \left( v(x) \right)^{m-1} v'(x) r(x)
 + \left( v(x) \right)^m r'(x)
\eq
and calculates the gcd
\bq
g(x) & = & \mbox{gcd}\left( u(x), u'(x) \right).
\eq
From eq. (\ref{squarefree1}) and eq. (\ref{squarefree2}) one sees that
$\left( v(x) \right)^{m-1}$ is a factor of the gcd and $u(x)$ can be 
written as
\bq
u(x) & = & \left( \frac{u(x)}{g(x)} \right) g(x).
\eq
Note that $g(x)$ divides $u(x)$ by the definition of the gcd.
This process can be repeated, until in each term all factors occur
with power one.
The computational cost is one (or several) gcd calculation(s).

\subsubsection{Berlekamp's algorithm}

After square-free decomposition we can assume that our
polynomial 
\bq
u(x) & = & p_1(x) p_2(x) ... p_k(x)
\eq
is a product of distinct primes.
We assume that $\mbox{deg}\;u(x) =n$.
Berlekamp's algorithm \cite{Berlekamp}
factorizes this polynomial in ${\mathbb Z}_p$,
where $p$ is a prime number.
For $0 \le k < n$ one obtains coefficients $q_{k,j}$ from the 
$\mbox{mod}\;u(x)$ representation of $x^{k p}$, e.g.
\bq
 x^{k p} & = & \left( q_{k,n-1} x^{n-1}
               + ... +
               q_{k,1} x + q_{k,0} \right) \; \mbox{mod} \; u(x).
\eq
This defines a $n \times n$ matrix
\bq
Q & = &
 \left(
 \begin{array}{cccc}
   q_{0,0} & q_{0,1} & ... & q_{0,n-1} \\
   ... &  & & ... \\
   q_{n-1,0} & q_{n-1,1} & ... & q_{n-1,n-1} \\
 \end{array}
 \right)
\eq
A non-trivial solution of
\bq
\label{berlekamp1}
\left( v_0, v_1, ..., v_{n-1} \right) Q & = & 
 \left( v_0, v_1, ..., v_{n-1} \right)
\eq
defines then a polynomial $v(x)$ by
\bq
v(x) & = & v_{n-1} x^{n-1} + ... + v_1 x + v_0.
\eq
Note that the solution of eq. (\ref{berlekamp1}) is obtained by
linear algebra in ${\mathbb Z}_p$.
The calculation of 
\bq
\mbox{gcd}\left( u(x), v(x) - s \right)
\eq
for $0 \le s < p$ will then detect the non-trivial factors of $u(x)$
in ${\mathbb Z}_p$.

\subsubsection{Hensel lifting}

Assume now that we have a factorization
\bq
u(x) & = & v_1(x) w_1(x)  \;\mbox{mod}\;p.
\eq
The Hensel lifting promotes this factorization to a factorization
\bq
u(x) & = & v_r(x) w_r(x)  \;\mbox{mod}\;p^r.
\eq
Since $u(x)$ was assumed to be square-free, we have 
$\mbox{gcd}(v_1(x),w_1(x)) = 1 \;\mbox{mod}\;p$.
For simplicity we also assume that the leading coefficient of $u(x)$
is $1$.
\\
Since $\mbox{gcd}(v_1(x),w_1(x)) = 1 \;\mbox{mod}\;p$
we can compute with the Euclidean algorithm polynomials
$a(x)$ and $b(x)$ with $\mbox{deg}\;a(x) < \mbox{deg}\;w_1(x)$
and $\mbox{deg}\;b(x) < \mbox{deg}\;v_1(x)$ such that
\bq
a(x) v_1(x) + b(x) w_1(x) & = & 1 \; \mbox{mod} \; p.
\eq
Suppose now that we are given $(v_r, w_r)$ and that we wish to compute
$(v_{r+1},w_{r+1})$.
To this aim one computes first a polynomial $c_r(x)$ such that
\bq
p^r c_r(x) & = & v_r(x) w_r(x) - u(x) \;\mbox{mod} \;p^{r+1}.
\eq
By polynomial division of $a(x) c_r(x)$ by $w_1(x)$ one obtains
a quotient $q_r(x)$ and remainder $a_r(x)$:
\bq
a(x) c_r(x) & = & q_r(x) w_1(x) + a_r(x) \;\mbox{mod}\;p.
\eq
One further sets
\bq
b_r(x) & = & b(x) c_r(x) + q_r(x) v_1(x)\;\mbox{mod}\;p.
\eq
$v_{r+1}$ and $w_{r+1}$ are now given by
\bq
v_{r+1} & = & v_r(x) - p^r b_r(x) \;\mbox{mod} \;p^{r+1}, \nonumber \\
w_{r+1} & = & w_r(x) - p^r a_r(x) \;\mbox{mod} \;p^{r+1}.
\eq
As an example let us consider $u(x) = x^2 + 27 x +176$.
This polynomial is already square-free and using Berlekamp's algorithm we try to
find a factorization in ${\mathbb Z}_3$. We use the asymmetric representation
$\{ 0,1,2\}$ for ${\mathbb Z}_3$ instead of the more symmetric representation
$\{ -1,0,1 \}$.
We find the factorization
\bq
u(x) & = & x^2 +2 \; \mbox{mod} \; 3 \nonumber \\
     & = & (x+1) (x+2)  \; \mbox{mod} \; 3.
\eq
Hensel lifting yields then
\bq
u(x) & = & (x+7) (x+2) \; \mbox{mod} \; 9 \nonumber \\
     & = & (x+16) (x+11) \; \mbox{mod} \; 27.
\eq
The factorization in ${\mathbb Z}_{27}$ agrees already with the factorization in ${\mathbb Z}$.

\subsection{Symbolic integration}

Textbook integration techniques are often not more than a collection 
of heuristic tricks, together with a look-up integration table.
A major achievement, which was initiated by the development of computer
algebra systems, was the invention of a systematic procedure to decide
whether a given function from a certain class of functions has an
integral inside this class.
If this is the case, the procedure
provides also a constructive method to calculate this integral.
This is the Risch integration algorithm 
\cite{Risch:1968}.
It is beyond the scope of these notes to describe this algorithm in 
all details, but some important features can already be seen, if we restrict
ourselves to rational functions.
\\
A rational function is a quotient of polynomials in the integration variable
$x$:
\bq
f(x) & = & \frac{a(x)}{b(x)} 
       =
           \frac{a_m x^m + ... + a_1 x + a_0}
                {b_n x^n + ... + b_1 x + b_0}.
\eq
Suppose that we know the factorization of the denominator over ${\mathbb C}$:
\bq
b_n x^n + ... + b_1 x + b_0 & = & 
 b_n (x-c_1)^{m_1} ... (x-c_r)^{m_r},
\eq
with $m_1+...+m_r=n$.
Then after polynomial division with remainder and partial fraction
decomposition we can write $f(x)$ in the form
\bq
\label{ratfct1}
f(x) & = & p(x) + \sum\limits_{i=1}^r \sum\limits_{j=1}^{m_i}
 \frac{d_{ij}}{(x-c_i)^j},
\eq
where $p(x)$ is a polynomial in $x$ and the $d_{ij}$ are complex numbers.
The integration of $p(x)$ is trivial. For the pole terms we have
\bq
\label{ratfct2}
\int \frac{d x}{(x-c_i)^j} & = &
 \left\{
 \begin{array}{cc}
   \ln \left( x- c_i \right), & j=1, \\
   \frac{1}{(1-j)} \frac{1}{(x-c_i)^{j-1}}, & j > 1. \\
 \end{array}
 \right.
\eq
The major inconvenience for this approach is the need to factor the denominator
completely and thereby introducing algebraic extensions (like square roots or
complex numbers), even if the integrand and the integral can entirely be
expressed in a smaller domain (like ${\mathbb Q}$).
As an example consider the integral
\bq
\label{ratfctex}
\int dx \frac{1-x^2}{(1+x^2)^2} & = & \frac{x}{1+x^2}.
\eq
The method above rewrites the integrand as 
\bq
\frac{1-x^2}{(1+x^2)^2} & = & 
 - \frac{1}{2} \frac{1}{(x+i)^2}
 - \frac{1}{2} \frac{1}{(x-i)^2},
\eq
and introduces the complex unit $i$, which cancels out in the final result.
\\
As can be seen from eq. (\ref{ratfct1}) and eq. (\ref{ratfct2})
the result for the integration of a rational function can always be written
as a rational function plus some logarithms.
It is desirable to compute as much of the integral
as possible in the domain of the integrand, and to find the minimal
algebraic extension necesary to express the integral.
This is done in two step: The first step uses Hermite's reduction method
and reduces the problem to one where the denominator is square-free.
The second step applies then
the Rothstein-Trager algorithm to obtain the logarithmic part.
We first consider Hermite's reduction method. Let
\bq
f(x) & = &
 p(x) + \frac{a(x)}{b(x)},
\eq
with $\mbox{deg}\;a(x) < \mbox{deg}\;b(x)$ and $\mbox{gcd}(a(x),b(x))=1$.
Assume further that $m$ is the highest power to which irreducible
factors occur in the factorization of the denominator, e.g.
\bq
b(x) & = & u(x) \left( v(x) \right)^m.
\eq
With the help of the Euclidean algorithm we may then calculate
polynomials $r(x)$ and $s(x)$ such that
\bq
r(x) u(x) v'(x) + s(x) v(x) & = & \frac{1}{1-m} a(x).
\eq
Then we obtain for the integral
\bq
\int dx \frac{a(x)}{u(x) \left( v(x) \right)^m}
 & = & 
 \frac{r(x)}{\left( v(x) \right)^{m-1}}
 + \int dx \frac{(1-m) s(x) - u(x) r'(x)}{u(x) \left( v(x) \right)^{m-1}},
\eq
and the problem is reduced to one with a smaller power of $v(x)$
in the denominator.
Therefore we can assume that
we are left with an integral of the form $f(x)=a(x)/b(x)$,
where $\mbox{deg}\;a(x) < \mbox{deg}\;b(x)$ and $b(x)$ is square-free.
This integral will yield a result of the form
\bq
\int dx \frac{a(x)}{b(x)} & = & 
 \sum\limits_{i=1}^n r_i \ln\left(x-c_i\right)
\eq
where the $c_i$'s are the zeros of $b(x)$  and the $r_i$'s are the residues
of $f(x)$ at the $c_i$'s.
There are efficient algorithms for the determination of the constants
$c_i$ and $r_i$.
The first algorithm was invented by Rothstein \cite{Rothstein} 
and Trager \cite{Trager}.
The method was later improved by Trager and Lazard and Rioboo \cite{Lazard}.
\\
\\
Before concluding this section, let us comment on the Risch integration
algorithm.
The integration method outlined above for the integration of 
rational functions was generalized by Risch
to an algorithm for elementary functions.
The class of elementary functions is defined as follows:
An elementary function of a variable $x$ is a function that can be obtained
from rational functions
by repeatedly adding a finite number of logarithms, exponentials
and algebraic numbers or functions (for example algebraic numbers like
the complex unit $i$ or square roots).
Any (finite) nested combination of those functions is again an elementary
function.
\\
For a systematic approach to integration it is useful to reformulate
some concepts in an algebraic way:
A differential field is a field $F$ of characteristic $0$ with a mapping
$D : F \rightarrow F$, which satisfies
\bq
D\left( f + g \right) & = & D\left(f\right) + D\left(g\right), \nonumber \\
D\left( f \cdot g \right) & = & D\left(f\right) \cdot g + f \cdot D\left(g\right).
\eq
The mapping $D$ is called a derivation or differential operator.
An example for a differential field is given by the field of rational
functions in one variable with $D$ being the usual differentiation.
\\
We have seen in example eq. (\ref{ratfctex}) that the integral for this 
example is again a rational function.
In general, however, the integral of a rational function is expressed
by logarithms and rational functions.
In this case, the logarithms are an extension to the original differential
field of rational functions.
In the case of elementary functions, we have to consider logarithmic,
exponential and algebraic extensions.
\\
Let $F$ be a differential field and let $G$ be a differential extension of it.
A function $g \in G$ is called logarithmic over $F$, if there exists a $f \in F$
such that
\bq
g' & = & \frac{f'}{f}.
\eq
A function $g \in G$ is called exponential over $F$, if there exists a $f \in F$
such that
\bq
\frac{g'}{g} & = & f'.
\eq
A function $g \in G$ is called algebraic over $F$, if there exists a polynomial $p \in F[z]$
such that
\bq
p(g) & = & 0.
\eq
Given an elementary function $f$, the Risch algorithm will decide whether
the integral of $f$ can be expressed as an elementary function.
If this is the case, the algorithm
provides also a constructive method 
to express the integral in terms
of elementary functions.
If this is not the case, we know at least, that the integral cannot be
expressed in terms of elementary functions.

\subsection{Symbolic summation}

Symbolic summation can be thought of as the discrete analog of 
symbolic integration.
For indefinite integration one looks for a function $F(x)$ such that
$F'(x) = f(x)$.
Then the definite integral over $f(x)$ is given by
\bq
\int\limits_a^b dx f(x) & = & F(b) - F(a).
\eq
For indefinite summation one looks for a function $T(k)$ defined for
integers $k$, such that the difference
\bq
\label{symsum1}
T(k+1) - T(k) & = & t(k)
\eq
equals a given function $t(k)$.
The definite sum is then given by
\bq
\sum\limits_{k=a}^{b-1} t(k) & = & T(b) - T(a)
\eq
One now looks for an algorithm, which for a given class of functions $f(k)$
decides, if the indefinite sum can be expressed in terms of functions
of this class. If the answer is positive, we would like to have a function
$T(k)$, which fullfills eq. (\ref{symsum1}). 
This problem has been solved for the class of ``hypergeometric terms''.
A sum of hypergeometric terms
\bq
\label{symsum3}
\sum\limits_{k = 0}^{n-1} t(k)
\eq
is a sum for which the ratio of two consecutive terms 
is a rational function $r(k)$ of the summation index $k$, e.g.
\bq
\label{symsum2}
\frac{t(k+1)}{t(k)} 
 & = & 
 r(k).
\eq
A function $t(k)$, which satisfies eq. (\ref{symsum2}) is called
a hypergeometric term.
In the following we will outline Gosper's algorithm \cite{Gosper}.
Suppose first that we can write $r(k)$ in the form
\bq
r(k) & = &
 \frac{a(k)}{b(k)} \frac{c(k+1)}{c(k)},
\eq
where $a(k)$, $b(k)$ and $c(k)$ are polynomials in $k$ and
\bq
\mbox{gcd}\left(a(k), b(k+j) \right) & = & 1,
\eq
for all non-negative integers $j$.
In fact, such a factorization exists for every rational function $r(k)$ 
and there
is a systematic way to find it.
We then look for a rational function $x(k)$, which satisfies
the first order difference equation
\bq
a(k) x(k+1) - b(k-1) x(k) & = & c(k).
\eq
If such a solution $x(k)$ does not exist, the sum in eq. (\ref{symsum3})
cannot be done within the class of hypergeometric terms.
Otherwise, the solution is given by
\bq
\sum\limits_{k = 0}^{n-1} t(k) & = &
 \frac{b(n-1) x(n)}{c(n)} t(n).
\eq
This completes Gosper's algorithm.
Up to now we considered indefinite summation. Although 
a solution for an indefinite summation problem may not exist,
we might be able to find a solution for a specific definite summation 
problem.
Definite summation problems have been studied for sums of the type
\bq
f(n) & = & \sum\limits_{k=-\infty}^\infty F(n,k),
\eq
where both
\bq
\frac{F(n+1,k)}{F(n,k)} \;\;\; \mbox{and} \;\;\;
\frac{F(n,k+1)}{F(n,k)}
\eq
are rational functions of $n$ and $k$.
Algorithms for these type of sums have been given by
Sister Celine, Wilf and Zeilberger and Petkovsek.
The book by Petkovsek, Wilf and Zeilberger \cite{Petkovsek:Book}
gives a good introduction to this subject.

\subsection{Gr\"obner bases}

In this paragraph we consider simplifications with the help
of Gr\"obner bases.
Assume that we have a (possibly rather long) 
expression $f$, which is a polynomial in
several variables $x_1$, ..., $x_k$.
In addition we have several siderelations of the form
\bq
s_j(x_1,...,x_k) & = & 0, \;\;\;1 \le j \le r,
\eq
which are also polynomials in $x_1$, ..., $x_k$.
A standard task is now to simplify $f$ with respect to the siderelations $s_j$,
e.g. to rewrite $f$ in the form
\bq
 f & = & a_1 s_1 + ... + a_r s_r + g,
\eq
where $g$ is ``simpler'' than $f$
The precise meaning of ``simpler'' requires the introduction of an order
relation on the multivariate polynomials.
As an example let us consider the expressions
\bq
f_1 = x + 2 y^3, & & 
f_2 = x^2,
\eq
which we would like to simplify with respect to the siderelations
\bq
s_1 & = & x^2 + 2 x y, \nonumber \\
s_2 & = & xy + 2 y^3 -1.
\eq
As an order relation we choose lexicographic ordering, e.g. $x$ is ``more
complicated'' as $y$, and $x^2$ is ``more complicated'' than $x$.
This definition will be made more precise below.
A naive approach would now take each siderelation, determine its ``most
complicated'' element, and replace each occurence of this element in the 
expression $f$ by the more simpler terms of the siderelation.
As an example let us consider for this approach the simplification
of $f_2$ with respect to the siderelations $s_1$ and $s_2$:
\bq
f_2 = x^2 = s_1 - 2 x y = s_1 - 2 y s_2 + 4 y^4 - 2y,
\eq
and $f_2$ would simplify to $4 y^4 - 2y$. 
In addition, since $f_1$ does not contain $x^2$ nor $x y$, the naive approach would not
simplify $f_1$ at all.
However, this is not the complete
story, since if $s_1$ and $s_2$ are siderelations, any linear
combination of those is again a valid siderelation.
In particular,
\bq
s_3 & = & y s_1 - x s_2 = x
\eq
is a siderelation which can be deduced from $s_1$ and $s_2$.
This implies that $f_2$ simplies to $0$ with respect to the siderelations
$s_1$ and $s_2$.
Clearly, some systematic approach is needed.
The appropriate tools are ideals in rings, and Gr\"obner bases for these
ideals.
\\
We consider multivariate polynomials in the ring $R[x_1,...,x_k]$.
Each element can be written as a sum of monomials of the form
\bq
c x_1^{m_1} ... x_k^{m_k}.
\eq
We define a lexicographic order of these terms by
\bq
c x_1^{m_1} ... x_k^{m_k} >
c' x_1^{m_1'} ... x_k^{m_k'},
\eq
if the leftmost nonzero entry in $(m_1-m_1', ..., m_k-m_k')$ is positive.
With this ordering we can write any element $f \in R[x_1,...,x_k]$ 
as
\bq
f & = & \sum_{i=0}^n h_i
\eq
where the $h_i$ are monomials and 
$h_{i+1} > h_i$ with respect to the lexicographic order.
The term $h_n$ is called the leading term and denoted $\mbox{lt}(f) = h_n$.
Let $ B=\{b_1,...,b_r\} \subset R[x_1,...,x_k]$ 
be a (finite) set of polynomials.
The set
\bq
\l B \r = \l b_1, ..., b_r \r & = & 
\left\{ \left. 
       \sum\limits_{i=1}^r a_i b_i \right| a_i \in R[x_1,...,x_k] \right\}
\eq
is called the ideal generated by the set $B$. The set $B$ is also called
a basis for this ideal.
(In general, given a ring $R$ and a subset $I \subset R$, $I$ 
is called an ideal
if $a+b \in I$ for all $a,b, \in I$ and $r a \in I$ for all $a \in I$ and
$r \in R$. Note the condition for the multiplication:
The multiplication has to be closed with respect
to elements from $R$ and not just $I$.)
\\
Suppose that we have an ideal $I$ and a finite subset $H \subset I$.
We denote by $\mbox{lt}(H)$ the set of leading terms of $H$ and, 
correspondingly by $\mbox{lt}(I)$ the set of leading terms of $I$.
Now suppose that the ideal generated by $\mbox{lt}(H)$ is identical
with the one generated by $\mbox{lt}(I)$, e.g. $\mbox{lt}(H)$ is a basis
for $\l \mbox{lt}(I) \r$.
Then a mathematical theorem guarantees that $H$ is also a basis for $I$, e.g.
\bq
\l \mbox{lt}(H) \r = \l \mbox{lt}(I) \r
 & \Rightarrow & 
 \l H \r = I
\eq
However, the converse is in general not true, e.g. if $H$ is a basis
for $I$ this does not imply that $\mbox{lt}(H)$ is a basis for
$\l \mbox{lt}(I) \r$.
A further theorem (due to Hilbert) states however that there exists
a subset $G \subset I$ such that 
\bq
\l G \r = I \;\;\;\mbox{and}\;\;\; \l \mbox{lt}(G) \r = \l \mbox{lt}(I) \r ,
\eq
e.g. $G$ is a basis for $I$ and $\mbox{lt}(G)$ is a basis for
$\l \mbox{lt}(I) \r$.
Such a set $G$ is called a Gr\"obner basis for $I$.
Buchberger \cite{Buchberger} gave an algorithm to compute $G$.
The importance of Gr\"obner bases for simplifications
stems from the following theorem:
Let $G$ be a Gr\"obner basis for an ideal $I \subset R[x_1,...,x_k]$ and $f \in R[x_1,...,x_k]$.
Then there is a unique polynomial $g \in R[x_1,...,x_k]$ with
\bq
f - g \in I
\eq
and no term of $g$ is divisible by any monomial in $\mbox{lt}(G)$.
\\
In plain text: $f$ is an expression which we would like to simplify
according to the siderelations defined by $I$.
This ideal is originally given 
by a set of polynomials 
$\{s_1,...,s_r\}$ and the siderelations are supposed to be of the
form $s_i=0$.
From this set of siderelations a Gr\"obner basis $\{b_1,...,b_{r'}\}$
for this ideal is calculated.
This is the natural basis for simplifying the expression $f$.
The result is the expression $g$, from which the  
``most complicated'' terms of $G$ have been eliminated, e.g. the terms $\mbox{lt}(G)$.
The precise meaning of ``most complicated'' terms depends on the
definition of the order relation.
\\
In our example, $\{s_1,s_2\}$ is not a Gr\"obner basis for
$\l s_1, s_2 \r$, since $\mbox{lt}(s_1)=x^2$ and $\mbox{lt}(s_2)=xy$
and
\bq
\mbox{lt}\left( y s_1 -x s_2 \right) = x \;\; \in\!\!\!\!\!/ \;\;
 \l \mbox{lt}(s_1), \mbox{lt}(s_2) \r.
\eq
A Gr\"obner basis for $\l s_1, s_2 \r$ is given by
\bq
\left\{ x, 2y^3 -1 \right\}.
\eq
With $b_1=x$ and $b_2=2 y^3 -1$ as a Gr\"obner basis, $f_1$ and $f_2$ can 
be simplified as follows:
\bq
f_1 & = & b_1 + b_2 + 1, \nonumber \\
f_2 & = & x b_1 + 0,
\eq
e.g. $f_1$ simplifies to $1$ and $f_2$ simplifies to $0$.

\subsection{Remarks}
\label{subsec:classes}

The Risch integration algorithm and Gosper's algorithm for summation both operate
on a class of functions and can decide if the integration- or summation problem
can be solved inside this class of functions.
If this is the case, they also provide the solution in terms of functions of this class.
The relevant classes of functions are the class of elementary functions for integration
and the class of hypergeometric terms for summation.
Unfortunately, these classes do not contain important functions relevant to loop 
calculations in perturbative quantum field theory.
For example, Euler's dilogarithm 
\bq
\mbox{Li}_2(x) & = & - \int\limits_0^x dt \frac{\ln(1-t)}{t}
\eq
is not in the class of elementary functions.
For the class of hypergeometric terms, the situation is even worse.
The harmonic numbers 
\bq
S_1(n) & = & \sum\limits_{j=1}^n \frac{1}{j}
\eq
are not included in the class of hypergeometric terms.
Therefore, logarithms like
\bq
\label{transoneloop}
-\ln(1-x) & = & \sum\limits_{i=1}^\infty \frac{x^i}{i}, \nonumber \\
\mbox{Li}_2(x) & = & \sum\limits_{i=1}^\infty \frac{x^i}{i^2}, \nonumber \\
\frac{1}{2} \ln^2(1-x) & = & \sum\limits_{i=1}^\infty \frac{x^i}{i} \sum\limits_{j=1}^{i-1}
 \frac{1}{j},
\eq
are not in the class of hypergeometric terms.
The three functions in eq. (\ref{transoneloop}) are the three basic transcendental functions
occuring in one-loop calculations.
Beyond one loop, there are additional functions.

\section{Algorithms for high energy physics}

In this section I discuss algorithms relevant to perturbative
calculations in high energy physics.
Topics include:
Generation of all contributing Feynman diagrams, 
contraction of indices and Dirac algebra gymnastic,
reduction of tensor loop integrals to scalar loop integrals
and the evaluation of scalar loop integrals in terms
of analytic functions.
This section can only offer a selection of topics.
Not included are for example methods based on Mellin transformations
\cite{Vermaseren:1998uu}
or asymptotic expansions. The last method has been reviewed in 
\cite{Harlander:1998dq,Steinhauser:2002rq}.

\subsection{Graph generation}

The first step in any perturbative calculation is usually the determination
of all relevant Feynman diagrams.
This is trivial for processes involving not more than a handful diagrams,
but requires a systematic procedure for processes involving a few hundreds
or more diagrams.
An efficient algorithm has been developed by Nogueira \cite{Nogueira:1993ex} and implemented into the
program QGRAF \cite{QGRAF}.
The algorithm avoids recursive generation of Feynman diagrams
and comparisons of stored diagrams.
The algorithm for the generation of Feynman graphs is divided into
two steps:
In the first step all relevant topologies are generated.
A topology is just a collection of nodes and edges, which
connect nodes.
A topology with $n$ nodes is represented by a $n \times n$
adjacency matrix $A$.
The entry $A_{ij}$ of the adjacency matrix
denotes the number of edges connecting the nodes
$i$ and $j$.
The algorithm generates all relevant adjacency matrices.
Using an order relation for adjacency matrices, one can avoid
to generate similar adjacency matrices, e.g. matrices which may be
related to each other by a permutation of the nodes.
In the second step the external nodes are labelled with the external
particles and the edges and internal nodes are substituted by propagators
and interaction vertices in all possible ways compatible with the Feynman
rules.
To avoid to generate equivalent Feynman graphs more than once during 
this stage, the symmetry group of each Feynman graph is determined
and a further order relation is used to return only the graph which
is the ``smallest'' within its symmetry class with respect to this order
relation.
In addition the symmetry factor is returned.
\\
Other algorithms for the generation of Feynman diagrams
have been considered for example by
K\"ublbeck, B\"ohm and Denner \cite{Kublbeck:1990xc} 
and implemented into FeynArts \cite{Hahn:2000kx} as well as
by Kajantie, Laine and Schr\"oder \cite{Kajantie:2001hv}.

\subsection{Contraction of indices and Dirac algebra}

After all diagrams have been generated, they are translated to mathematical
expressions with the help of Feynman rules.
Edges correspond to propagators and vertices to interaction vertices.
As an example for Feynman rules we give the rule for gluon propagator 
in a covariant gauge and the rule for 
the quark-gluon vertex:
\begin{eqnarray}
\begin{picture}(50,20)(0,10)
 \Gluon(0,15)(50,15){4}{4}
\end{picture} & = &
 \frac{-i}{k^2} \left(g_{\mu \nu}-(1-\xi)\frac{k_{\mu}k_{\nu}}{k^{2}}\right) \delta_{a b},
 \nonumber \\
\begin{picture}(50,30)(0,15)
 \Gluon(0,20)(30,20){4}{2}
 \Vertex(30,20){2}
 \ArrowLine(30,20)(50,40)
 \ArrowLine(50,0)(30,20)
\end{picture} & = &
 i g \gamma^{\mu} T^a_{ij}. \nonumber \\
 & & 
\end{eqnarray}
The resulting expressions involve summations over repeated indices.
The contraction of these indices can be done by applying succesivly a few
rules:
\bq
\label{contract}
& &
g_{\mu \nu} g^{\nu \rho} = g_\mu^{\;\;\; \rho},
\;\;\;
g_{\mu \nu} p^\nu = p_\mu,
\;\;\;
g_{\mu \nu} \gamma^\nu = \gamma_\mu,
\;\;\;
g_{\mu}^{\;\;\; \mu} = D,
 \nonumber \\
& &
p_\mu q^\mu = p q,
\;\;\;
p_\mu \gamma^\mu = p\!\!\!/,
\;\;\;
\gamma_\mu \gamma^\mu = D.
\eq
If a contraction over a string over Dirac matrices occurs, like
in 
\bq
\gamma_\mu \gamma_\nu \gamma^\mu,
\eq
one uses first the anti-commuation relations of the Dirac matrices
\bq
\{ \gamma_{\mu}, \gamma_{\nu} \} & = & 2 g_{\mu \nu} 
\eq
to bring the two matrices with the same index next to each other and uses
then eq. (\ref{contract}).
For the calculation of squared amplitudes there will be always a trace over
the strings of Dirac matrices, which can be evaluated according to
\bq
\label{tracegamma}
\mbox{Tr}\; {\bf 1} & = & 4, \\
\mbox{Tr}\; \gamma_{\mu_1} \gamma_{\mu_2} ... \gamma_{\mu_{2n}}
 & = & g_{\mu_1 \mu_2} \mbox{Tr}\; \gamma_{\mu_3} ... \gamma_{\mu_{2n}}
     - g_{\mu_1 \mu_3} \mbox{Tr}\; \gamma_{\mu_2} \gamma_{\mu_4} ... \gamma_{\mu_{2n}}
     + ...
     + g_{\mu_1 \mu_{2n}} \mbox{Tr}\; \gamma_{\mu_2} ... \gamma_{\mu_{2n-1}}.
 \nonumber 
\eq
Traces over an odd number of Dirac matrices vansih.
The recursive procedure for the evaluation of a trace over $2n$ Dirac matrices
will generate $(2n-1)(2n-3) ... 3 \cdot 1 = (2n-1)!!$ terms.
This number grows exponentially with $n$.
If there are no further relations between the indices $\mu_1$, ..., $\mu_{2n}$
this is indeed the number of terms in the final result.
However, in almost all practical applications, the free indices
$\mu_1$, ..., $\mu_{2n}$ get contracted with a tensor, which is at least
symmetric in some of its indices.
In that case the recursive procedure is inefficient.
A better way consists in splitting the string of Dirac matrices
into two parts 
\cite{Kennedy:1981kp}:
\bq
\gamma_{\mu_1} \gamma_{\mu_2} ... \gamma_{\mu_{2n}}
 & = & 
 \left( \gamma_{\mu_1} \gamma_{\mu_2} ... \gamma_{\mu_j} \right)
 \left( \gamma_{\mu_{j+1}} ... \gamma_{\mu_{2n}} \right)
\eq
This process is repeated until each factor contains only a few Dirac
matrices (say not more than two).
One then rewrites each factor in terms of the basis
\bq
\Gamma^{(0)} & = & 1, \nonumber \\
\Gamma^{(1)}_{\nu} & = & \gamma_\nu, \nonumber \\
\Gamma^{(2)}_{\nu_1 \nu_2} & = & \frac{1}{2} \left[ \gamma_{\nu_1}, \gamma_{\nu_2}\right], \;\;\; \mbox{etc.}
\eq
The general term of this basis has the form
\bq
\Gamma^{(j)}_{\nu_1 \nu_2 ... \nu_j} & = & 
 \gamma_{\left[ \nu_1 \right.} \gamma_{\nu_2} ... \gamma_{\left. \nu_j \right]},
\eq
where $[ ... ]$ denotes anti-symmetrization.
Obviously,
\bq
\mbox{Tr}\; \Gamma^{(j)}_{\nu_1 \nu_2 ... \nu_j} & = & 0 \;\;\;\mbox{for $j \ge 1$}.
\eq
Products of these basis elements can be combined with a Clebsch-Gordan
type formula:
\bq
\Gamma^{(i)}_{\mu_1 ... \mu_i} \Gamma^{(j)}_{\nu_1 ... \nu_j}
 & = & 
\sum\limits_{k=|i-j|}^{i+j} \frac{1}{k!} 
 C_{\mu_1 ... \mu_i;\nu_1 ... \nu_j}^{\rho_1 ... \rho_k}
 \Gamma^{(k)}_{\rho_1 ... \rho_k}.
\eq
The Clebsch-Gordan coefficients 
$C_{\mu_1 ... \mu_i;\nu_1 ... \nu_j}^{\rho_1 ... \rho_k}$ 
are anti-symmetric in the sets
$\{\mu_1 ... \mu_i\}$, $\{\nu_1 ... \nu_j\}$ and $\{\rho_1 ... \rho_k\}$.
Consider now the situation where
the tensor with which the free indices $\mu_1$, ..., $\mu_{2n}$
are contracted, is symmetric in $\mu_i$ and $\mu_j$.
In this situation, any term involving
a Clebsch-Gordan coefficient with $\mu_i$ and $\mu_j$ in the same
index field can immediately be discarded.
This can lead to a considerable speed-up.
\\
\\
Minor complications occur if $\gamma_5$ appears in the calculation.
Within dimensional regularisation this requires a consistent definition of $\gamma_5$.
One possible definition is the 't Hooft - Veltman scheme, which takes $\gamma_5$ as a generic
four-dimensional object. As a consequence one has to distinguish between four-dimensional
and $D=4-2\eps$ dimensional quantities.
In this scheme $D$ is assumed to be greater than $4$.
It is further assumed that the
four-dimensional subspace and the $(-2\eps)$ dimensional subspace are
orthogonal.
For the calculation
one splits all quantities into a four-dimensional part (denoted with a tilde)
and a $(-2\eps)$ dimensional part (denoted with a hat).
\bq
 g_{\mu \nu} = \tilde{g}_{\mu \nu} + \hat{g}_{\mu \nu},
 & &
 \gamma_{\mu} = \tilde{\gamma}_{\mu} + \hat{\gamma}_{\mu}.
\eq
These quantities satisfy relations like
\bq
 \tilde{g}_{\mu}^{\;\;\;\mu} = 4 ,
 \;\;\;
 \hat{g}_{\mu}^{\;\;\;\mu} = -2 \varepsilon ,
 \;\;\;
 \tilde{g}_{\mu \nu} \hat{g}^{\nu \rho} = 0.
\eq
$\gamma_5$ is then defined as a generic four-dimensional object:
\bq
 \gamma_{5} = \frac{i}{4!} \varepsilon_{\alpha \beta \gamma \delta}
 \tilde{\gamma}^{\alpha} \tilde{\gamma}^{\beta}
 \tilde{\gamma}^{\gamma} \tilde{\gamma}^{\delta}.
\eq
As a consequence, $\gamma_5$ anti-commutes with the four-dimensional
Dirac matrices, but commutes with the remaining ones:
\bq
 \{ \gamma_{5}, \tilde{\gamma}_{\mu} \} = 0, 
 & &
\left[ \gamma_{5} , \hat{\gamma}_{\mu} \right] = 0.
\eq
The program FORM \cite{FORM} 
is one of the most efficient tools for manipulations involving contraction of indices
and traces over Dirac matrices.
The program TRACER \cite{Jamin:1993dp} has been written for the manipulations of the Dirac
algebra in the 't Hooft - Veltman scheme.

\subsection{One-loop integrals and Passarino-Veltman reduction}

We now consider the reduction of tensor loop integrals 
(e.g. integrals, where the loop momentum appears in the numerator)
to a set of scalar loop integrals (e.g. integrals, where the numerator
is independent of the loop momentum).
For one-loop integrals a systematic algorithm has been first worked
out by Passarino and Veltman \cite{Passarino:1979jh}.
Consider the following three-point integral
\begin{eqnarray}
I_3^{\mu\nu} & = & \int \frac{d^{D}k}{i \pi^{D/2}}
\frac{k^\mu k^\nu}{k^2 (k-p_1)^2 (k-p_1-p_2)^2},
\end{eqnarray}
where $p_1$ and $p_2$ denote the external momenta.
The reduction technique according to Passarino and Veltman consists in writing $I_3^{\mu\nu}$
in the most general form
in terms of form factors times external momenta and/or the metric tensor. In our example above
we would write
\begin{eqnarray}
\label{passarino}
I_3^{\mu\nu} & = & p_1^\mu p_1^\nu C_{21} + p_2^\mu p_2^\nu C_{22}
+ \{p_1^\mu, p_2^\nu\} C_{23} + g^{\mu\nu} C_{24},
\end{eqnarray}
where $\{p_1^\mu, p_2^\nu\} = p_1^\mu p_2^\nu + p_2^\mu p_1^\nu$.
One then solves for the form factors $C_{21}, C_{22}, C_{23}$ and $C_{24}$ by
first contracting both sides with the external momenta $p_1^\mu p_1^\nu$, $p_2^\mu p_2^\nu$,
$\{p_1^\mu, p_2^\nu \}$ and the metric tensor $g^{\mu\nu}$.
On the left-hand side the resulting scalar products between the loop momentum $k^\mu$ and the external
momenta are rewritten in terms of the propagators, as for example
\begin{eqnarray}
2 p_1 \cdot k & = & k^2 - (k-p_1)^2 + p_1^2.
\end{eqnarray}
The first two terms of the right-hand side above cancel propagators, whereas the last term does not involve the 
loop momentum anymore.
The remaining step is to solve for the formfactors $C_{2i}$ by inverting the matrix which one obtains on the 
right-hand side of
equation (\ref{passarino}).
Due to this step Gram determinants usually appear in the denominator of the final expression.
In the example above we would encounter the Gram determinant of the triangle
\begin{eqnarray}
\Delta_3 & = & 4 \left|
\begin{array}{cc}
p_1^2 & p_1\cdot p_2 \\
p_1 \cdot p_2 & p_2^2 \\
\end{array} \right|.
\end{eqnarray}
One drawback of this algorithm is closely related to these determinants : In a phase space
region where $p_1$ becomes collinear to $p_2$, the Gram determinant will tend to zero, and the 
form factors will take large values, with possible large cancellations among them. This makes
it difficult to set up a stable numerical program for automated evaluation of tensor loop
integrals.
There are modifications of the Passarino-Veltman algorithm, which avoid
to a large extent the appearance of Gram determinants.
These improved algorithms are based on spinor methods 
\cite{Pittau:1997ez}.

\subsection{Beyond one-loop}

The Passarino-Veltman algorithm is based on the observation, that for one-loop
integrals a scalar product
of the loop momentum with an external momentum can be expressed
as a combination of inverse propagators.
This property does no longer hold if one goes to two or more loops.
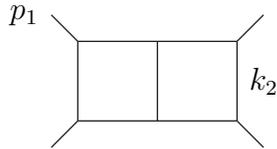
\begin{figure}
\begin{center}
\begin{picture}(100,70)(0,0)
\Line(20,20)(80,20)
\Line(20,50)(80,50)
\Line(20,20)(20,50)
\Line(50,20)(50,50)
\Line(80,20)(80,50)
\Line(10,10)(20,20)
\Line(80,20)(90,10)
\Line(20,50)(10,60)
\Line(80,50)(90,60)
\Text(5,60)[r]{$p_1$}
\Text(85,35)[l]{$k_2$}
\end{picture}
\caption{\label{irredscalprod} \it An example for irreducible scalar products in the numerator: The scalar product $2p_1 k_2$ cannot be expressed in terms of
inverse propagators.}
\end{center}
\end{figure}
Fig. (\ref{irredscalprod}) shows a two-loop diagram, for which the scalar 
product of a loop momentum with an external momentum cannot be expressed
in terms of inverse propagators.
\\
A different and more general method for the reduction of tensor integrals 
is needed.
In the following I will outline algorithms, which work beyond one loop.
The first step is to relate tensor integrals to scalar integrals with raised
powers of the propagators and different values of the space-time dimension $D$
\cite{Tarasov:1996br}.
The starting point is the
Schwinger parameterization for the propagators 
\begin{eqnarray}
\frac{1}{(-k^2)^\nu} & = & \frac{1}{\Gamma(\nu)} \int\limits_0^\infty dx \;
x^{\nu-1} \exp(x k^2).
\end{eqnarray}
Let us first consider a scalar two-loop integral.
Combining the exponentials arising from different propagators one
obtains a quadratic form in the loop momenta. 
For instance, for a given two-loop integral with loop momenta $k_1$ and $k_2$, 
one has then 
\begin{eqnarray}
I(D,\nu_1,\ldots,\nu_k)&=&
\int \frac{d^{D }k_1}{i\pi^{D /2}}
\int \frac{d^{D }k_2}{i\pi^{D /2}}
\frac{1}{(-k_1^2)^{\nu_1}\ldots (-k_n^2)^{\nu_n}} \\
&=&
\int \frac{d^{D }k_1}{i\pi^{D /2}}
\int \frac{d^{D }k_2}{i\pi^{D /2}} \left( \prod_{i=1}^n \frac{1}{\Gamma(\nu_i)}
\int^{\infty}_0 dx_i\, x_i^{\nu_i-1} \right) \exp\left ( \sum_{i=1}^n x_i k_i^2 \right), 
 \nonumber 
\end{eqnarray}
and
\begin{equation}
\label{eq:quadraticform}
\sum_{i=1}^n x_i k_i^2 = a\, k_1^2 + b\, k_2^2 + 2\, c \,k_1 \cdot k_2 + 2\, d
\cdot k_1 + 2 \, e \cdot k_2 + f\ .
\end{equation}
The momenta $k_3,\ldots,k_k$ are 
linear combinations of the loop momenta $k_1,k_2$ and the external momenta. 
The coefficients $a$, $b$, $c$, $d^\mu$, $e^\mu$ and $f$ are directly 
readable from the actual graph: $a (b) = \sum x_i$, where the sum 
runs over the legs in the $k_1$ $(k_2)$ loop, and $c = \sum x_i$ with 
the sum running over
the legs common to both loops. 
With a suitable change of variables for the loop momenta $k_1, k_2$, one can 
diagonalize the quadratic form 
and the momentum integration can be performed as Gaussian integrals 
over the shifted loop momenta according to
\begin{eqnarray}
\int \frac{d^{D }k}{i\pi^{D /2}} \exp \left( {\cal P} k^2 \right) & = & 
 \frac{1}{{\cal P}^{D /2}}.
\end{eqnarray}
Let us now consider a tensor integral. After the change of variables for the diagonalization
of the quadratic form, we have a polynomial in the Schwinger parameters and the loop momenta
$k_1$ and $k_2$ in the numerator.
Integrals with an odd power of a loop momentum in the numerator vanish by symmetry, while
integrals with an even power of the loop momentum can be related by Lorentz
invariance to scalar integrals:
\begin{eqnarray}
\int \frac{d^{D }k}{i\pi^{D /2}} k^\mu k^\nu f(k^2) & = & 
 \frac{1}{D } g^{\mu\nu} \int \frac{d^{D }k}{i\pi^{D /2}} k^2 f(k^2), \\
\int \frac{d^{D }k}{i\pi^{D /2}} k^\mu k^\nu k^\rho k^\sigma f(k^2) & = & 
 \frac{1}{D (D +2)} 
  \left( g^{\mu\nu} g^{\rho\sigma} + g^{\mu\rho} g^{\nu\sigma} + g^{\mu\sigma} g^{\nu\rho} \right) 
  \int \frac{d^{D }k}{i\pi^{D /2}} k^2 f(k^2). \nonumber
\end{eqnarray}
The generalization to arbitrary higher tensor structures is obvious.
In the remaining Schwinger parameter integrals, the tensor integrals 
introduce additional factors of the parameters $x_i$ and of $1/{\cal P}$.
These additional factors can be absorbed into scalar integrals with 
higher powers of propagators and shifted dimensions, by introducing 
operators ${\bf i}^+$,  which raise the power of propagator
$i$ by one, or an operator ${\bf d}^+$ that increases the dimension by two,
\begin{eqnarray}
& &
\frac{1}{\Gamma(\nu_i)} \int\limits_0^\infty dx_i \; x_i^{\nu_i-1} x_i \exp(x_i k^2)
 =  
 \nu_i \frac{1}{\left(-k_i^2\right)^{\nu_i+1}}
   =
 \nu_i {\bf i}^+ \frac{1}{\left(-k_i^2\right)^{\nu_i}},
 \nonumber \\
& &
 \frac{1}{{\cal P}} \frac{1}{{\cal P}^{D /2}}
 = 
 \int \frac{d^{(D +2)}k}{i\pi^{(D +2)/2}} \exp \left( {\cal P} k^2 \right)
 =
 {\bf d}^+ \int \frac{d^{D }k}{i\pi^{D /2}} \exp \left( {\cal P} k^2 \right).
\end{eqnarray}
All Schwinger integrals are rewritten in terms of these scalar integrals.
Therefore, using an intermediate Schwinger parametrization, we have expressed
all tensor integrals in terms of scalar integrals.
The price we paid is that these scalar integrals involve higher powers of the propagators
and/or have shifted dimensions.
Each integral can be specified by its topology, its value for the dimension $D$ and 
a set of indices, denoting the powers of the propagators.
In general the number of different integrals is quite large.
There are now two different methods to proceed:
The first method 
(integration by parts and solutions of differential equations)
observes that there are non-trivial relations among the obtained 
set of integrals and employs an elimination procedure to reduce this set to a small
set of so-called master integrals.
These master integrals are then evaluated explicitly.
\\
Since some integrals have to be evaluated explicitly anyway, one may ask if there
is an efficient algorithm which evaluates directly these integrals for arbitrary
$D$ and arbitrary set of indices. This is the philosophy of the second method
(expansion of transcendental functions).

\subsubsection{Integration by parts and solutions of differential equations}

Let us start with the method, which reduces the (large) number of
scalar integrals to a (small) set of master integrals.
There exist non-trivial relations among various scalar integrals.
Some of these relations can be obtained from integration-by-part identities
\cite{Chetyrkin:1981qh}
or the invariance of scalar integrals under Lorentz transformations 
\cite{Gehrmann:1999as}.
Integration-by-part identities are based on the fact, that the integral
of a total derivative is zero:
\bq
\int \frac{d^D k}{i \pi^{D/2}} 
\frac{\partial}{\partial k^{\mu}} v^{\mu} f(k,p_i) & = & 0.
\eq
Here, $k$ is the loop momentum, the $p_i$'s are the external momenta
and $v$ can either be a loop momentum or an external momentum.
Working out the derivative yields a relation among
several scalar integrals.
In a similar way one obtains relations based on Lorentz-invariance.
A scalar integral is evidently invariant under an infinitessimal
Lorentz transformation, parametrized as
\bq
p^{\mu} \to p^{\mu} + \delta p^{\mu} = 
p^{\mu} + \delta \epsilon^{\mu}_{\nu} p^{\nu} \qquad 
\mbox{with} \qquad \delta \epsilon^{\mu}_{\nu} = - \delta
\epsilon^{\nu}_{\mu}\;.
\eq
This implies that
\bq
\left(p_1^{\nu}\frac{\partial}{\partial
    p_{1\mu}} - p_1^{\mu}\frac{\partial}{\partial
    p_{1\nu}} + \ldots + p_n^{\nu}\frac{\partial}{\partial
    p_{n\mu}} - p_n^{\mu}\frac{\partial}{\partial
    p_{n\nu}}\right) I(p_1,\ldots,p_n) = 0 \;,
\eq
where $I(p_1,...,p_n)$ is a scalar integral with the external
momenta $p_i$.
Again, working out the derivatives yields a relation among
several scalar integrals.
\\
Each relation is linear in the scalar integrals and in principle
one could use Gauss elimination to reduce the set of scalar integrals
to a small set of master integrals.
In practice this approach is inefficient.
An efficient algorithm has been given by Laporta \cite{Laporta:2001dd}.
The starting point is to introduce an order relation for
scalar integrals. This can be done in several ways, a possible
choice is to order the topologies first: A scalar integral corresponding
to a topology $T_1$ is considered to be ``smaller'' than an integral
with topology $T_2$, if $T_1$ can be obtained from $T_2$ by pinching
of some propagators.
Within each topology, the scalar integrals can be ordered
according to the powers of the propagators and the dimension of space-time.
Laporta's algorithm is based on the fact that starting
from a specific topology, integration-by-part and Lorentz-invariance
relations only generate relations involving this topology
and ``smaller'' ones.
To avoid to substitute a specific identity into a large number of other
identities, one starts from the ``smallest'' topology and generates all
relevant relations for this topology.
Inside this class, integrals with higher powers of the propagators
or higher dimensions are then expressed in terms of a few master integrals.
These manipulations involve only a small subset of the complete
system of integrals and relations and can therefore be done
efficiently.
Once this topology is completed, one moves on to the next
topology, until all topologies have been considered.
All integrals, which cannot be eliminated by this procedure
are called master integrals.
These master integrals have to be evaluated explicitly.
\\
To evaluate these integrals the
procedure used in 
\cite{Gehrmann:1999as,Gehrmann:2000zt} 
consists in finding first 
for each master integral a differential
equation, which this master integral has to satisfy.
The derivative is taken with respect to an external scale, or a
ratio of two scales.
It turns out that the resulting differential equations
are linear, inhomogeneous first order equations 
of the form
\bq
\frac{\partial }{\partial y} T(y) + f(y) T(y) = g(y),
\eq
where $y$ is usually a ratio of two kinematical invariants and 
$T(y)$ is the master integral under consideration.
The inhomogeneous term $g(y)$ is usually a combination of simpler master integrals.
In general, a first order linear inhomogeneous differential
equation is solved by first considering the corresponding homogeneous
equation.
One introduces an integrating factor 
\bq
M(y) = e^{\int f(y) d y},  
\eq
such that  $T_0(y)=1/M(y)$ solves the homogenous differential equation
($g(y)=0$). This 
yields the general solution of the inhomogenous equation as
\bq
T(y) = \frac{1}{M(y)} \left( \int g(y) M(y) d y + C\right),
\eq
where the integration constant $C$ can be adjusted to match the boundary 
conditions. 
This method yields a master integral in the form of transcendental
functions (e.g. for example hypergeometric functions),
which still have to be expanded in the small parameter $\eps$ of dimensional
regularization.
Although this can be done systematically and will be 
discussed in the next paragraph,
the authors of
\cite{Gehrmann:1999as,Gehrmann:2000zt}
followed a different road.
As an example let us consider the calculation for $e^+ e^- \rightarrow \mbox{3 jets}$, which involves three scales. This scales can be taken to be
$s_{12}$, $s_{23}$ and $s_{123}$. It is convenient to have only one dimension-full quantity $s_{123}$ and to introduce
two
dimensionless quantities $x_1=s_{12}/s_{123}$ and $x_2=s_{23}/s_{123}$.
Factoring out a trivial dimension-full normalization factor, one
writes 
down an ansatz for the solution of the differential equation 
as a Laurent expression in $\eps$.
Each term in this Laurent series is a sum of terms, consisting of
basis functions times
some unknown (and to be determined) coefficients.
This ansatz is inserted into the differential equation and the unknown 
coefficients
are determined order by order from the differential equation.
This apporach will succeed if we know in advance the right set
of basis functions.
The basis functions are a subset of multiple polylogarithms.
In sec. (\ref{sec:shuffling}) we already defined in eq. (\ref{Gfunc})
the functions
$G(z_1,...,z_k;y)$.
We now sligthly enlarge this set and define
$G(0,...,0;y)$ with $k$ zeros for $z_1$ to $z_k$ to be
\bq
G(0,...,0;y) & = & \frac{1}{k!} \left( \ln y \right)^k.
\eq
This permits us to allow trailing zeros in the sequence
$(z_1,...,z_k)$.
We can now define two subsets of these functions.
The first subset are harmonic polylogarithms \cite{Remiddi:1999ew} 
for which $y=x_1$ and all
$z_i$ are from the set $\{0,1\}$.
The second subset are two-dimensional harmonic polylogarithms 
\cite{Gehrmann:2000zt} for
which $y=x_2$ and all
$z_i$ are from the set $\{0,x_1,1-x_1,1\}$.
The ansatz consists in taking the harmonic polylogarithms and the 
two-dimensional harmonic polylogarithms as a set of basis functions
for the three-scale problem $e^+ e^- \rightarrow \mbox{3 jets}$.
The coefficients for harmonic polylogarithms are rational functions
in $x_1$ and $x_2$, the coefficients of the two-dimensional harmomic
polylogarithms may in addition also contain
(one-dimensional) harmonic polylogarithms.

\subsubsection{Expansion of transcendental functions}

In this paragraph we now discuss the second approach. Instead of reducing
first a large set of scalar integrals to a small set of master integrals,
which have to be worked out explicitly, we may ask if there is an efficient
way to evaluate the scalar integrals directly
\cite{Moch:2001zr,Weinzierl:2002hv,Moch:2002hm}.
The basic observation is that many relevant integrals can be expressed 
as (possibly nested) sums involving Gamma-functions.
As an example let us discuss
the one-loop triangle with two external masses defined by
\begin{eqnarray}
     \mbox{Tri}_2(m,\nu_1,\nu_2,\nu_3;x_1) = 
         \left( -s_{123} \right)^{-m+\varepsilon+\nu_{123}}
         \int \frac{d^D k_1}{i \pi^{D /2}}
         \frac{1}{\left(-k_1^2\right)^{\nu_1}}
         \frac{1}{\left(-k_2^2\right)^{\nu_2}}
         \frac{1}{\left(-k_3^2\right)^{\nu_3}},
\end{eqnarray}
where $k_2=k_1-p_1-p_2$, $k_3=k_2-p_3$ and $x_1=s_{12}/s_{123}$.  
We inserted a prefactor $\left( -s_{123} \right)^{-m+\varepsilon+\nu_{123}}$
to make the integral dimensionless and used the short-hand notation
$\nu_{ij}=\nu_i+\nu_j$ for sums of indices.
Within this approach the integral is needed for arbitrary (integer) powers
of the propagators and possibly also in shifted dimensions $6-2\eps$,
$8-2\eps$, etc..
The space-time dimension is written as $D=2 m - 2\eps$, where $m$ is an
integer.
It is not too hard to derive a series representation for this integral,
consisting 
of a 
combination of hypergeometric functions $_2F_1$.
\begin{eqnarray}
\lefteqn{
         \mbox{Tri}_2(m,\nu_1,\nu_2,\nu_3;x_1) \,= \,
             \frac{ 
                     \Gamma(\varepsilon-m+\nu_{23})
                     \Gamma(1-\varepsilon+m-\nu_{23})
                     \Gamma(m-\varepsilon-\nu_{13})
                  }{
                     \Gamma(\nu_1) \Gamma(\nu_2) \Gamma(\nu_3) 
                     \Gamma(2m-2\varepsilon-\nu_{123})
                   } 
}
\\ & & 
             \times 
             \sum\limits_{n=0}^\infty
                 \frac{x_1^{n}}{n!} 
             \left[
             x_1^{m-\varepsilon-\nu_{23}} 
             \frac{
                     \Gamma(n+\nu_1) 
                     \Gamma(n-\varepsilon+m-\nu_2) 
                  }{
                     \Gamma(n+1+m-\varepsilon-\nu_{23}) 
                   }
                 - 
             \frac{
                     \Gamma(n+\nu_3) 
                     \Gamma(n-m+\varepsilon+\nu_{123})
                  }{
                     \Gamma(n+1-m+\varepsilon+\nu_{23}) 
                   }
             \right]\, .
\nonumber 
\end{eqnarray}
One observes that the small paramter $\eps$ of dimensional
regularization appears in the Gamma-functions.
This expression has now to be expanded into a Laurent series
in $\eps$.
For this particular example it is possible to convert the hypergeometric
functions
into an integral representation, to expand the integrand and to perform
the resulting integrals.
However, more complicated topologies have a representation as nested sums,
for which a useful integral representation is not known.
Fortunately, the expansion in $\eps$ can be done systematically at the
level of nested sums.
Using the formula
\bq
\Gamma(x+1) & = & x \; \Gamma(x),
\eq
all Gamma functions can be synchronized to the form $\Gamma(n+a \eps)$.
They are then expanded using the formula
\bq
\label{expgamma}
\lefteqn{
\Gamma(n+\eps) = \Gamma(1+\eps) \Gamma(n) } & & \nonumber \\
& & \times \left( 1 + \eps Z_1(n-1) + \eps^2 Z_{11}(n-1) + \eps^3 Z_{111}(n-1) + ... 
+ \eps^{n-1} Z_{11...1}(n-1) \right). \nonumber \\
\eq
Here Euler-Zagier sums, introduced in sec. (\ref{sec:shuffling}) 
in eq. (\ref{EulerZagier}) appear.
Collecting terms for each order in $\eps$, we obtain expressions involving
products of Euler-Zagier sums with the same upper summation limit.
Since Euler-Zagier sums form an algebra, the multiplication can be done.
After some additional simple manipulations
(partial fraction decomposition and adjusting of summation limits)
we obtain terms of the form
\bq
\sum\limits_{n=1}^\infty \frac{x_1^n}{n^{m_0}} Z_{m_1,...,m_k}(n-1).
\eq
Since these are exactly the harmonic polylogarithms \cite{Remiddi:1999ew}
\bq
\label{harmpolylog}
H_{m_0,...,m_k}(x_1) & = & \sum\limits_{n=1}^\infty \frac{x_1^n}{n^{m_0}} Z_{m_1,...,m_k}(n-1),
\eq
we are already done.
For topologies involving more scales, we have to generalize this approach.
The basic quantities are no longer Euler-Zagier sums, but Z-sums 
defined by
\bq 
  Z(n;m_1,...,m_k;x_1,...,x_k) & = & \sum\limits_{n\ge i_1>i_2>\ldots>i_k>0}
     \frac{x_1^{i_1}}{{i_1}^{m_1}}\ldots \frac{x_k^{i_k}}{{i_k}^{m_k}}.
\eq
Clearly, for $x_1=...x_k=1$ we recover the Euler-Zagier sums.
All algorithms work also for this generalization, in particular, there is 
a multiplication formula.
Of particular importance are the sums to infinity:
\bq
\label{Gonch}
\mbox{Li}_{m_k,...,m_1}(x_k,...,x_1) & = &
 Z(\infty;m_1,...,m_k;x_1,...,x_k)
\eq
These are called multiple polylogarithms.
There is a close relation between the functions $G(z_1,...,z_k;y)$ and
$\mbox{Li}_{m_k,...,m_1}(x_k,...,x_1)$.
Let us denote by $G_{m_1,...,m_k}(z_1,...,z_k;y)$ the $G$-function, where $m_1-1$ zeros
preceed
$z_1$, $m_2-1$ zeros are inserted between $z_1$ and $z_2$ etc..
For example $G_{2,3,1}(z_1,z_2,z_3;y) = G(0,z_1,0,0,z_2,z_3;y)$.
Then
\bq
G_{m_1,...,m_k}(z_1,...,z_k;y) & = & 
 (-1)^k \; \mbox{Li}_{m_k,...,m_1}\left(\frac{z_{k-1}}{z_k},...,\frac{z_1}{z_2},\frac{y}{z_1}\right), \nonumber \\
\mbox{Li}_{m_k,...,m_1}(x_k,...,x_1)
& = & (-1)^k G_{m_1,...,m_k}\left( \frac{1}{x_1}, \frac{1}{x_1 x_2}, ..., \frac{1}{x_1...x_k};1 \right),
\eq
showing that the iterated integral representation eq. (\ref{Gfunc})
and the representation as nested sums
eq. (\ref{Gonch})
describe the same class of functions.

\subsubsection{The cut technique}

We conclude the discussion of techniques for loop calculations
with a method based on unitarity \cite{Bern:1995cg}.
I will discuss this method for one-loop amplitudes.
It has been applied to a two-loop calculation by Bern et al. \cite{Bern:2000dn}.
\\
Within the unitarity based approach one chooses first a basis of integral functions $I_i \in {\cal F}$.
For one-loop calculations in massless QCD
a possible set consists of scalar boxes, triangles
and two-point functions.
The loop amplitude $A^{loop}$ is written as a linear combination of these functions
\begin{eqnarray}
\label{cutbased}
A^{loop} & = & \sum\limits_i c_i I_i + r.
\end{eqnarray}
The unknown coefficients $c_i$ are to be determined. $r$ is a rational
function in the invariants, not proportional to any element of the basis
of integral functions.
The integral functions themselves are combinations of rational functions,
logarithms, logarithms squared and dilogarithms.
The latter three can develop imaginary parts in certain regions
of phase space, for example
\bq
\mbox{Im} \ln \left( \frac{-s-i 0}{-t-i 0} \right) & = & - \pi \left[ \theta(s) -\theta(t) \right], \nonumber \\
\mbox{Im} \; \mbox{Li}_2 \left( 1 - \frac{(-s-i 0)}{(-t-i 0)} \right)
 & = & - \ln \left( 1 - \frac{s}{t} \right)
        \mbox{Im} \ln \left( \frac{-s-i 0}{-t-i 0} \right).
\eq
Knowing the imaginary parts, one can reconstruct uniquely the corresponding
integral functions.
In general there will be imaginary parts corresponding to different
channels (e.g. to the different possibilities to cut a one-loop
diagram into two parts).
The imaginary part in one channel of a one-loop amplitude 
can be obtained via unitarity from a phase space integral over
two tree-level amplitudes.
With the help of the Cutkosky rules we have
\begin{eqnarray}
\label{cutconstr}
\mbox{Im} \; A^{loop} & = & \mbox{Im} \int \frac{d^D k}{(2 \pi)^D} \frac{1}{k_1^2 + i 0}
\frac{1}{k_2^2 + i 0} A_L^{tree} A_R^{tree}.
\end{eqnarray}
$A^{loop}$ is the one-loop amplitude under consideration, $A^{tree}_L$ and
$A^{tree}_R$ are tree-level amplitudes appearing on the left and right side
of the cut in a given channel.
Lifting eq. (\ref{cutconstr}) one obtains
\begin{eqnarray}
A^{loop} & = & \int \frac{d^D k}{(2 \pi)^D} \frac{1}{k_1^2 + i 0}
\frac{1}{k_2^2 + i 0} A_L^{tree} A_R^{tree} + \;\mbox{cut free pieces},
\end{eqnarray}
where ``cut free pieces'' denote contributions which do not develop an imaginary
part in this particular channel.
By evaluating the cut, one determines the coefficients $c_i$ of the integral
functions, which have an imaginary part in this channel.
Iterating over all possible cuts, one finds all coefficients $c_i$.
\\
One advantage of a cut-based calculation is that one starts with tree amplitudes on both sides of the cut, which are already sums
of Feynman diagrams. 
Therefore cancellations and simplifications, which usually occur between
various diagrams, can already be performed before we start the calculation
of the loop amplitude.
\\
\\
In general, a cut-based calculation leaves as ambiguity the ration piece $r$
in eq. (\ref{cutbased}), which can not be obtained with this technique.
One example for such an ambiguity would be
\begin{eqnarray}
\int \frac{d^D k}{(2 \pi)^D} \frac{k^\mu k^\nu - \frac{1}{3} q^\mu q^\nu + \frac{1}{12} g^{\mu\nu} q^2}{k^2 (k-q)^2}.
\end{eqnarray}
This term does not have a cut and will therefore not be detected in a cut-based calculation. 
However, Bern, Dixon, Dunbar and Kosower \cite{Bern:1995cg}
have proven the following power counting criterion:
If a loop-amplitude has in some gauge a representation, in which all $n$-point loop integrals have at most
$n-2$ powers of the loop momentum in the numerator (with the exception of two-point integrals, which are allowed
to have one power of the loop momentum in the numerator), then the loop amplitude is uniquely determined
by its cuts. 
This does not mean that the amplitude has no cut-free pieces, but rather that all cut-free
pieces are associated with some integral functions.\\
In particular $N=4$ supersymmetric amplitudes satisfy the power-counting criterion above and are therefore
cut-constructible.
\\
\\
QCD does in general not satisfy the power-counting theorem and leaves
as an ambiguity the rational function $r$.
In principle one can obtain the rational piece $r$ by calculating higher order
terms in $\eps$ within the cut-based method.
At one-loop order an arbitrary scale $\mu^{2\varepsilon}$ is introduced in order to keep the coupling
dimensionless. In a massless theory the factor $\mu^{2\varepsilon}$ is always accompanied
by some kinematical invariant $s^{-\varepsilon}$ for dimensional reasons.
If we write symbolically
\begin{eqnarray}
A^{loop} & = & \frac{1}{\varepsilon^2} c_2 \left( \frac{s_2}{\mu^2} \right)^{-\varepsilon} 
+ \frac{1}{\varepsilon} c_1 \left( \frac{s_1}{\mu^2} \right)^{-\varepsilon}
+ c_0 \left( \frac{s_0}{\mu^2} \right)^{-\varepsilon} ,
\end{eqnarray}
the cut-free pieces $c_0 (s_0/\mu^2)^{-\varepsilon}$ can be detected at order $\varepsilon$:
\begin{eqnarray}
c_0 \left( \frac{s_0}{\mu^2} \right)^{-\varepsilon} & = & c_0 - \varepsilon c_0 \ln \left(\frac{s_0}{\mu^2}\right) + O(\varepsilon^2).
\end{eqnarray}
In practice, it is more efficient to take into account additional constraints,
like the factorization in collinear limits \cite{Bern:1994zx}, to determine the rational piece
$r$.

\subsection{Remarks}

Let us summarize the state-of-the-art of systematic algorithms for the computation
of loop integrals occuring in perturbative quantum field theory.
In sect. \ref{sec:classical} we discussed Risch's algorithm for symbolic integration and
Gosper's algorithm for symbolic summation.
These algorithms operate on a certain class of functions and provide the solution if it exists
within this class of functions.
The two classes of functions are the elementary functions for Risch's algorithm and 
the hypergeometric terms for Gosper's algorithm.
None of these two classes is sufficiently large to contain all functions, which occur
in the calculation of loop integrals.
A counter-example has been given in sect. \ref{subsec:classes}
with the three transcendental functions (dilogarithm, logarithm and logarithm squared)
which occur in one-loop calculations.
\\
Beyond one-loop, additional transcendental functions occur in loop calculations.
From the explicit two-loop calculations we now know that the class of functions
occuring in loop calculations, should contain the multiple polylogarithms.
There are now systematic algorithms, which allow the calculation of specific
loop integrals.
As examples we discussed an algorithm based on differential equations and an algorithm
based on the expansion of transcendental functions.
The former is based on the manipulation of iterated integrals, whereas the latter
is based on the manipulation of nested sums.
Although the algorithms are quite different, their output is within the same class
of functions.
\\
At the end of these lectures, 
let us look towards the future:
Consider the class of rational functions extended by the multiple polylogarithms.
Is there an algorithm, similar to Risch's or Gosper's algorithm, which decides,
if a given integral or sum can be expressed inside this class ?
If so, is there a constructive way to compute the answer ?
These are open questions.
Calculations in perturbative quantum field theory would certainly profit
from a positive answer to these questions.

\end{document}